# FROM THE BERLIN "ENTWURF" FIELD EQUATIONS TO THE EINSTEIN TENSOR III: March 1916

Galina Weinstein

*Visitor Scholar, Center for Einstein Studies, Department of Philosophy, Boston University*

*January 25, 2012*

I discuss Albert Einstein's 1916 General Theory of Relativity.
I show that in Einstein's 1916 review paper, "the Foundation of the General Theory of Relativity", he derived his November 25, 1915 field equations with an additional term on the right hand side involving the trace of the energy-momentum tensor (not in their generally covariant form, but he posed the condition, $\sqrt{-g} = 1$) using the equations he presented on November 4, 1915. Series of papers: Final paper.

## 1 Mid December to Mid January 1915: Exchange of Letters between Einstein and Ehrenfest

Einstein used to travel on a regular basis to Paul Ehrenfest, his Jewish physicist friend, who worked in the University of Leiden, and to Lorentz who also lived in the Netherlands; but he also corresponded with both of his best friends.

Already in 1907 Ehrenfest posed to Einstein the first query about the theory of special relativity. Ehrenfest brought Einstein to rethink the foundations of his then new theory of relativity, and Einstein's reply to Ehrenfest's query was important for the demarcation between a theory of relativity and Lorentz's ether-based theory.[1]

In 1909 Ehrenfest was not satisfied again in his paper, "Gleichförmige Rotation Starrer Körper und Relativitätstheorie".[2] The issue now was Lorentz contraction and rigid bodies that cannot really exist.

Suppose we are given a rigid cylinder of radius $R$. It is rotating with constant rotation around its axis. An observer at rest with respect to the cylinder measures its radius as $R'$. When the cylinder is moving, the observer at rest measures $R' < R$ due to the contraction of lengths. This is so because each element of the peripheral surface of the cylinder moves with instantaneous velocity $R'\omega$. But this seems to be wrong, because each element of the radius of the cylinder is perpendicular to the direction of the velocity of the cylinder. Thus $R' = R$. How can $R' < R$ and at the same time $R' = R$?[3]

Einstein published nothing directly on this question during the next few years, but his first published reference to the rigidly rotating disk occured in the first of two papers on static gravitational fields from February 1912.[4]

In his paper "Lichtgeschwindigkeit und Statik des Gravitationsfeldes" (the Speed of Light and the Statics of the Gravitational Fields"), Einstein referred to the rotating disc in section §1. He considered a system *K* with coordinates *x, y, z* in a state of uniform acceleration in the direction of its *x*-coordinate, and referred it to a non-accelerated system; the acceleration of *K*'s origin possesses no velocity, and it is constant. "According to the equivalence hypothesis this system *K* is strictly equivalent to a system at rest in which there exists a certain kind of mass-free static gravitational field". Einstein determines that the special measurements of *K* are performed by means of measuring rods. When these are compared with one another in a state of rest at the same location in *K* they possess the same length. Einstein thus assumes that the laws of geometry hold for the lengths so measured, and for the relations between the coordinates and for other lengths. But this stipulation is not always permitted, because it contains physical assumptions that might after all be wrong. They do not hold for a uniformly rotating system. If our definition were applied, then owing to the Lorentz contraction, the ratio of the circumference to the diameter would have to be different from $\pi$. The measuring rod and the coordinate axes are to be conceived as rigid bodies. Einstein says that this is permitted even though according to the theory of relativity *the rigid body cannot really exist*. The reason is that we can replace the measuring rigid body by a great number of non-rigid bodies arranged in a row.[5]

In 1912 emission theories flourished. Ehrenfest again posed a somewhat annoying query. Ehrenfest published a paper comparing Einstein's views on light propagation with those of Walter Ritz. [6] Ehrenfest noted that although both approaches involved a particulate description of light – Einstein invoked the quanta of light and Ritz proposed particles of light – nevertheless, Ritz's theory constituted a "real" emission theory (in the Newtonian sense), while Einstein's theory was more akin to the ether conception; since it postulated that the velocity of light is independent of the velocity of its source. Ehrenfest suggested possible experiments to distinguish between the two theories and noted the necessary of carrying out such empirical test. Einstein reacted to Ehrenfest's above paper by writing Ehrenfest, "I was not annoyed in the least by your article! On the contrary, such considerations are quite familiar to me from the pre-relativistic time".[7]

By this time Einstein was probably annoyed a little by his friend's queries, although he loved him very much; much later in 1934 Einstein wrote in memoriam to Ehrenfest, "Added to this was the increasing difficulty of adaptation to new thoughts which always confronts the man past fifty. I do not know how many readers of these lines will be capable of fully grasping that tragedy. Yet it was this that primarily occasioned his escape from life".[8]

In December 1915 Ehrenfest posed a new query; this time Ehrenfest asked Einstein about the Hole Argument from section §12 of his 1914 review article. This query was probably the most annoying one of all the queries that Ehrenfest had asked Einstein over the years. The reason was that Einstein silently dropped the Hole Argument during October 1915 and he did not mention it in his November 1915 papers.

Stachel wrote that Einstein had to explain just what is wrong with the Hole Argument against general covariance, in which he believed so strongly for over two years. He explained to Ehrenfest that in section §12 of his paper of "last year, everything is correct (in the first 3 paragraphs) up to the italicized part at the end of the third paragraph". Einstein then told Ehrenfest that in the Hole Argument G(x) and G'(x) represented two different gravitational fields with respect to the same reference system. The correction should be that the reference system has no meaning and that the realization of two gravitational fields in the same region of the continuum is impossible. Einstein wrote Ehrenfest, "The following consideration should replace §12", and that was the Point Coincidence Argument, later presented in a new 1916 review paper to be presented and analyzed in this chapter. Einstein also gave his friend Besso the same explanation he had given a week earlier to Ehrenfest.[9]

What was wrong with his Hole Argument could be seen from Einstein's explanation on December 14 to Moritz Schlick that, through the general covariance of the field equations, "time and space lose the last remnant of physical reality. All that remains is that the world is to be conceived as a four-dimensional (hyperbolic) continuum of 4 dimensions".[10]

Ehrenfest soon posed to Einstein more queries, and in an exchange of letters Einstein gradually and patiently answered them all; and while answering the queries, Einstein developed the basis of a new review article. Finally Ehrenfest's queries led Einstein to reconsider his November 4 1915 field equations, and to add another term to these equations, and by thus to rederive the November 25 field equations for a coordinate system, with respect to which $\sqrt{-g} = 1$ holds everywhere. This was the first formulation of the General Theory of Relativity.

In March 1916 Einstein sent to the editor of the *Annalen der Physik*, Wilhelm Wien, a new review article on the general theory of relativity. Einstein wrote the paper on 20/3/1916.[11] The paper was published two months later, in May 1916. This paper was a correction of the 1914 review article "Die formale Grundlage der allgemeinen Relativitätstheorie",[12] in light of the November 1915 papers. However, the new 1916 paper was actually more than that: it presented new procedures and derivations.

In the 1916 paper Einstein considered the field equations valid for matter-free gravitational field, $G_{\mu\nu} = 0$, which he had already written in the November 18 paper. These equations were now valid for a coordinate system in such a way that with respect to this system, $\sqrt{-g} = 1$ holds everywhere. Einstein followed the methods of his November 4 paper. He corrected his November 4 paper by which he was led to the November 25 field equations that were valid for a coordinate system with respect to which $\sqrt{-g} = 1$.

**2 The 1916 Review Article: "The Foundation of the General Theory of Relativity".**

There are two versions of the 1916 review paper: the first is a 46 pages manuscript from 1916 that is found in the Einstein Archives, and the second is a published version from the *Annalen der Physik*.[13] The manuscript was edited when published in the *Annalen*. For instance, Einstein wrote on page 11 of the manuscript "dieses §" – meaning, this section (Einstein used to designate sections of his papers by §). And it was edited to "dieses paragraphen".[14]

There is still a third version of the paper… In 1915 Sommerfeld edited a book *Das Relativitatsprinzip*, a collection of the path breaking papers of special and general relativity (with comments by Sommerfeld). The book began with an abridged version of Lorentz's 1904 paper on the theory of electron; then afterwards, were brought Einstein's 1905 two relativity papers, and his 1911 paper, and another paper on relativity by Lorentz. [15] Sommerfeld asked Einstein to contribute his 1914 review "Entwurf" paper to the book, but the later refused. The book was afterwards republished with new editions, and in later editions it included a *corrected version* of the 1916 *Annalen der Physik* paper presenting the generally covariant general theory of relativity. This version is somewhat different from the original 1916 *Annalen* review article, and thus the English translation of this paper by W. Perrett and G.B. Jeffery from the English equivalent book (1952), *is not* the English translation of the original 1916 paper!

Below Einstein's 1916 theory is analyzed using both sources: the 1916 *Annalen* paper and the manuscript. However, keep in mind that what you read in the *Annalen* is not always Einstein's own language, because the paper was edited.

**2.1 The Special Theory of Relativity**

In the manuscript of the 1916 review article, Einstein crossed out after the title of the paper two sub-titles that he intended to give straight at the beginning of the paper:

"A. Prinzipielle Erwägungen zum Postulat der Relativität"

(Fundamental Considerations on the Postulate of Relativity)

And

"§1. Die Spezielle Relativitätstheorie"[16]

Einstein decided first to write an introduction and then he brought part A with the same title and afterwards section §1 with somewhat different title, although it is related to the special theory of relativity. Einstein wrote straight at the beginning of his manuscript "special relativity", because his principle of equivalence was a local principle of equivalence, and the flat Minkowski metric was always his starting point.

Einstein started the introduction of the printed version of the 1916 review article by saying, "The theory outlined in what follows is the most far-reaching possible generalization of the now generally accepted theory, called 'theory of relativity'; I present the latter in the following, distinguishing it from the former 'special relativity', and assume it is already known".[17] He presented a new theory, and called it "allgemeinen Relativitätatheorie". The November 1915 papers did not yet present a full-fledged theory. And now for the first time Einstein lays down a *theory* which is based on his 1915 papers.

The paper begins with section "A Fundamental Considerations of the Postulate of Rrelativity", section §1"Comments on the special theory of relativity".[18] This section begins with qualitative explanations of Mach's ideas, general covariance, the equivalence principle, the general principle of relativity, and the Point Coincidence Argument. It resembles Einstein's 1905-1912 papers: thought experiments and friendly explanations of principles.

Einstein started his paper with a short claim that his theory departed from special relativity with regards to the postulate of the constancy of the velocity of light. He first defined the "special principle of relativity", which is also satisfied by the mechanics of Galileo and Newton:

"If a coordinate system K is chosen so that, in relation to it, the laws of physics hold in their simplest form, the *same* laws are valid with respect to any other coordinate system K' moving in uniform translation with respect to K".[19]

Einstein called this postulate special because it defines the case "when K' has a motion of *uniform translation* compared to K, but that the equivalence of K' and K does not extend to the case of *non-uniform* motion of K' relatively to K".[20]

Einstein then noted, "Thus the special theory of relativity does not depart from classical mechanics through the postulate of relativity, but only by the postulate of the constancy of the velocity of light in vacuum, from which, in combination with the special principle of relativity, the relativity of simultaneity, the Lorentz transformation, and the related laws for the behavior of moving rigid bodies and clocks, follow, in the well known way".[21]

In this and other respects Einstein later did not consider special relativity as revolutionary. Carl Seelig wrote, "As opposed to several interpreters, Einstein would not agree that the relativity theory was a revolutionary event. He used to say: 'In the [special] relativity theory it is no question of a revolutionary act but of a natural development of lines which have been followed for centuries' ".[22]

Einstein also did not consider special relativity as revolutionary because,[23]

"The modification to which the special theory of relativity has subjected the theory of space and time is indeed profound, but *an* important point has remained unaffected. Also as requested by the special theory of relativity, namely, the laws of geometry are directly interpreted as laws relating to the possible relative positions (at rest) of solid bodies, and, more generally, the laws of kinematics are to be interpreted as laws which describe the relations of measuring bodies and clocks. To two selected material points of a stationary (rigid) body there always corresponds a distance of quite definite length, which is independent of the location and orientation of the body, as well as of the time; to two selected positions of the hands of a clock at rest relatively to the (legitimate) reference system, there always corresponds an interval of time of a definite length, which is independent of place and time".

Einstein added, "It will soon be seen that the general theory of relativity cannot adhere to this simple physical interpretation of space and time".[24]

But before demonstrating via the Disk thought experiment why this simple picture was untenable in the General Theory of Relativity, Einstein gave yet another reason for why the special theory of relativity should be extended. Special relativity dealt with symmetric reference systems, that is, inertial reference systems. In order to extend the "special principle of relativity", one needed to consider *non-symmetric* reference systems, i.e., non-inertial reference systems.

In a 1914 *Scientia* paper Einstein proposed a thought experiment, originally suggested by Newton in the *Principia*,[25] the two globes thought experiment.[26] He considered two masses, and did not say anything about their shape. It was implicitly assumed that these masses were *symmetric*, and the problem was whether the Newtonian explanation applied, or rather the Machian one. In section §2 of the 1916 paper Einstein adapted this 1914 thought experiment, in order to extend the "special principle of relativity", and he presented the two masses as *non*-symmetric masses. This way special relativity no more applied to the two non-symmetric masses (one inertial and the other non-inertial).

**2.2 The Two Globes Thought Experiment and Mach's Ideas**

Let us come back to the first part of the *Scientia* 1914 globes thought experiment:
"In outer space there are two masses floating at a great distance from all celestial bodies. The masses are close enough to each other to be able to exert mutual influence. An observer follows the motions of both bodies by constantly sighting in the direction of the line connecting the two masses toward the vault of the fixed stars.

He will assume that the line of sight traces a closed line on the visible vault of the fixed stars, which does not change its position with respect to the visible vault of the fixed stars".[27]

The 1916 masses are fluid masses of the same size and nature. They hover freely at so great distance from each other that, one takes into consideration only the gravitational forces that arise from the interaction of different parts of the same body. They are measured by observers, which are at rest with respect to the bodies; an observer who is at rest on each mass judges the other mass as rotating with constant angular velocity. [28]

Einstein added to the 1914 description that the masses were "fluid", and this way he could present *apparently* non-symmetric two bodies of different shapes, but of the same size and nature: "Now, we consider that the surfaces of both bodies ($S_1$ and $S_2$) are measured by means of measuring rods (relatively at rest), and it follows that the surface of $S_1$ is a sphere, and that of $S_2$ is an ellipsoid of revolution.[29]

This is the first part of the thought experiment. The first part of the two globes thought experiment appears to be an extension of the magnet and conductor thought experiment from Einstein's relativity paper of 1905. Recall Einstein's own words from 1921: [30] "Then [1907] there came to me the happiest thought of my life in the following form: The gravitational field is considered in the same way and has only a relative existence like the electric field generated by magneto-electric induction".

After presenting the 1905 magnet and conductor thought experiment in 1905 Einstein wrote, "Examples of this sort, […] lead to the conjecture that the phenomena of electrodynamics as well as those of mechanics possess no properties corresponding to the idea of absolute rest".[31]And the globes thought experiment was intended to demonstrate that this could be extended to accelerated motions and to the theory of gravitation using Mach's principle (still not defined as a principle).

Einstein thus solved the problem with the two apparently *non*-symmetric fluid masses $S_1$ and $S_2$ in much the same way as he had done with the above magnet and conductor thought experiment. He was guided by Mach's ideas and discussed the asymmetry problem:

We ask: What is the reason that body $S_1$ behaves differently than body $S_2$? [32] Newtonian mechanics does not give a satisfactory answer to this question. The laws of mechanics apply to the space $R_1$, in respect to which the body $S_1$ is at rest, but not to the space $R_2$, in respect to which the body $S_2$ is at rest. But the legitimate space $R_1$ of Galileo, thus introduced, is a "*merely factitious* cause", and not a thing that can be observed. It is also clear that Newton's mechanics demands that the factitious cause $R_1$ is responsible for the behavior of the bodies $S_1$ and $S_2$.

A satisfactory answer to the question raised can only be the following one: the physical system consisting of $S_1$ and $S_2$ reveals within itself no imaginable cause to which the differing behavior of $S_1$ and $S_2$ can be referred. The cause must therefore lie *outside* this system. We have to admit that the general laws of motion, which in particular determine the shapes of $S_1$ and $S_2$, must be such that the mechanical behavior of $S_1$ and $S_2$ is also conditioned, in quite essential respects, by distant masses which we have not included in the system under consideration. [33]

Immediately after the two globes thought experiment Einstein formulated the following version of the principle of general relativity,[34]

"*The laws of physics must be of such a nature that they apply to systems of reference in any kind of motion*. Along this road we arrive at an extension of the postulate of relativity".

**2.3 The Equivalence Principle**

Einstein then returned to the coordinate-dependent description from the beginning of his review article of 1914. He considered the two systems he had presented in 1914: one K, which is in uniform translation motion, and is Galilei-Newtonian coordinate system, and the other K', which is in uniform rotation relative to K. Centrifugal forces then act on the masses at rest relative to K', while they do not act upon the masses which are at rest relative to K. [35]

He would soon extend this case, but first he was occupied with an elaboration of the equivalence principle he had formulated in his Prague works from 1911-1912:

K' is moving with uniformly translated acceleration with respect to K. Relative to K a mass is moving with uniform motion in a straight line. Relative to K' a mass would have an accelerated motion such that its acceleration and direction of acceleration are independent of the material composition and physical state of the mass.

Einstein then presented the equivalence principle,

"Can an observer at rest relatively to K' infer that he is on a 'really' accelerated reference system? The answer to this question is negative; because the above-mentioned behavior of the freely moving masses relative to K' can be equally interpreted in the following way. The reference system K' is unaccelarated; but in the considered space-time regions there is a gravitational field, which generates the accelerated motion of the bodies with respect to K'." [36]

**2.4 Modification of the Principle of the Constancy of the Velocity of Light**

After presenting the Equivalence principle, Einstein replied to his critics such as Abraham, Nordsrtöm, and others, "Similarly, it is obvious from experience that the principle of the constancy of the velocity of light in vacuum must be modified. Since we easily recognize that the path of a ray of light with respect to K' must in general be curved, if with respect to K light is propagated in a straight line with a certain constant velocity".[37]

Already in his 1907 paper for the *Yearbook for Radioactivity and Electronics*, *Jahrbuch der Radioaktivität und Elektronik*, Einstein recognized that according to his then principle of equivalence *Aequivalenzprinzip*, which he formulated in that paper in section §17:

$c(1 + \gamma\xi/c^2) = c(1 + \phi/c^2)$. [38]

This equation later led Einstein to conclude that the velocity of light in a gravitational field is a function of the place. In 1907 Einstein reasoned: "It follows from this that the light rays, that do not move along the $\xi$-axis, are bent by the gravitational field; the change of direction amounts to, as can be easily seen, per centimeter light path $\gamma/c^2 \sin \varphi$, where $\varphi$ is the angle between the direction of gravity and that of the light ray".[39]

In his Prague paper of 1911, "Uber den Einfluβ der Schwerkraft auf die Ausbreitung des Lichtes" ("On the Influence of Gravitation on the Propagation of Light"). Einstein began this paper by saying, "It is clear namely, that rays of light, passing close to the sun, experience by its gravitational field the same deflection which follows from the theory here to be brought forward, so that the apparent increase in the angular distance occurring between the sun and a fixed star appearing near to it amounts to nearly a second of arc".[40] Einstein concluded that "*a ray of light passing near the sun would undergo a deflection of amount* $4 \cdot 10^{-6}$ *= 0,83 arc seconds*".[41]

Like its predecessor for the static gravitational field from 1911, the "Entwurf" theory predicted the same value for the deflection of light in a gravitational field of the sun, 0.83 seconds of arc.[42] The November 18 paper corrected this value and Einstein reproduced the November 18 derivation at the end of his 1916 review paper.

Over the years Abraham, Nordström, and others attacked Einstein's theory of gravitation for not following the line of special relativity; especially they felt that Einstein changed the postulate of the constancy of the velocity of light. Abraham said: "Already a year ago, A. Einstein has given up the essential postulate of the constancy of the speed of light by accepting the effect of the gravitational potential on the speed of light, in his earlier theory.[43] And Nordström explained that Einstein's hypothesis that the speed of light *c* depends on the gravitational potential led to considerable problems such as revealed in the Einstein-Abraham dispute.[44] The bending of light predicted by Einstein's theory (and still not verified experimentally) seemed to be in clash with this postulate.

Einstein clarified in his 1916 review article the demarcation between the special and general theories of relativity, and the relation between them. Light rays that moved in straight lines signified an affiliation with Euclidean geometry.

**2.5 The Disk Thought Experiment**

In section §3 Einstein dealt with the disk experiment. His first mention of the rotating disk in print was in his 1912 paper dealing with the static gravitational fields of 1912; and after the 1912 paper, the rotating-disk argument occurred in Einstein's writings only in the 1916 review article.[45] Einstein did not mention the rotating disk problem in any of his papers on gravitation theory from 1907 through 1915.[46]

The initial motivation for presenting the rotating disk thought experiment in 1916 was to show that coordinates of space and time have no direct physical meaning; since coordinates have no direct physical meaning Euclidean Geometry breaks down.

Einstein explained that in classical mechanics and in special relativity, space-time measurements of coordinates ($x_1$, $x_2$, $x_3$, $x_4$) are done with rods and standard clocks. With these we define lengths and times in all inertial reference frames. The notions of coordinates and measurements in classical mechanics and in special relativity presuppose the validity of Euclidean geometry.

Einstein now considered the two systems, the Galilean system K, and the other K', which is in uniform rotation relative to K. He then showed by a thought experiment that we are unable to define properly coordinates in K', and Euclidean geometry breaks down for K'. He concluded that we are also unable to properly define time by clocks at rest in K' either. Therefore, the coordinates of space and time have no direct physical meaning with respect to K'. This realization brought Einstein to general covariance, which he formulated after presenting the Disk thought experiment.

In 1916 Einstein again considered the two systems of reference, the Galilean K and the one K', which is in uniform rotation relative to K. The origin of both systems, as well as their axes of Z, permanently coincide one with another. The *circle* around the origin in the X, Y plane of K is regarded at the same time as a *circle* in the X', Y' plane of K'. The Disk is a circle. "We now imagine that the circumference and diameter of this circle are measured with a unit measure (infinitely small relative to the radius), and we form the quotient of the two results".[47]

If the experiment is performed with a measuring rod at rest relative to a Galilean system K, the quotient will be π. With a measuring rod at rest relative to K', the quotient will be greater than π.[48] Einstein added, "This can be seen easily, if the whole process of measurement is viewed from the system at 'rest' K, taking into consideration that the periphery undergoes a Lorentz contraction, while the measuring rod applied to the radius does not".[49] It follows that therefore the lengths measurements have no direct meaning and Euclidean geometry does not apply to K'.[50]

After propounding on lengths measurements of the 1916 *circle*, Einstein was discussing time measurements. Einstein did not discuss this matter in his 1912 paper, "The Speed of Light and the Statics of the Gravitational Fields", presenting the first version of the disk experiment. [51]

Einstein was now able to extend his 1907 and 1911 descriptions of clocks running at different rates to the explanation of the rate of clocks in the rotating disk story.

In section §18 of his 1907 paper on the theory of relativity for the *Yearbook for Radioactivity and Electronics*, Einstein considered a reference system S' that is uniformly accelerated relative to a non-accelerated system S in the direction of its X-axis. The clocks of S' are set at time t' and Einstein asked what is the rate of the clocks in the next time element $\tau$?[52] If the totality of readings of the clocks of S' is the "local time" $\sigma$ of the system S', then,

$\sigma = \tau(1 + \gamma\xi/c^2)$.

According to the equivalence principle, this equation is also valid for a coordinate system in which a homogeneous gravitational field is acting. In that case Einstein set $\phi = \gamma\xi$, where, $\phi$ is the gravitational potential, and obtained:

$\sigma = \tau(1 + \phi/c^2)$.[53]

In section §19 Einstein used the above equation. If a clock indicating $\sigma$ is located in a point P of a gravitational potential $\phi$, then according to the above equation, its reading will be $(1 + \phi/c^2)$ times greater than the time $\tau$. It runs $(1 + \phi/c^2)$ times faster than an identical clock located at the coordinate origin. For an observer located somewhere in space, the clock in point P runs $(1 + \phi/c^2)$ times faster than the clock at the coordinate origin.[54]

In his Prague paper of 1911, Einstein reconsidered the measurement of time in uniform accelerated systems, "If we measure time in $S_1$ with a clock U, *we must measure the time in* $S_2$ *with a clock that goes* $1 + \Phi/c^2$ *slower than the clock U if you compare it with the clock U in the same place*".[55]

In the 1916 paper Einstein imagined two clocks of identical constitution placed, one at the origin of coordinates, and the other at the periphery of the circle. Both clocks are observed from the system at "rest" K. "According to a known result from special relativity – judged from K – the clock at the periphery of the circle goes more slowly than the other clock at the origin, because the clock at the former [the circumference] is in motion and the latter [at the origin] is at rest".[56]

An observer who is located at the origin, and who is capable of observing the clock at the circumference by means of light, would be able to see the periphery clock lagging behind the clock beside him. He will interpret this observation as showing that the clock at the periphery "really" goes more slowly than the clock at the origin. He will thus define time in such a way that the rate of the clock depends upon its location.[57]

In summary, when we measure the circumference of the circle of K' from the system K, then "the measuring-rod applied to the periphery undergoes a Lorentzian contraction, while the one applied along the radius does not"; and when we require measurement of time events in K', then judged from K, "the clock at the periphery of the circle goes more slowly than the other clock at the origin".

Einstein concluded after presenting the disk story that "In the general theory of relativity, space and time cannot be defined in such a way that spatial coordinate differences be directly measured by the unit measuring rod, and time by a standard clock".[58] This conclusion signifies that Euclidean geometry breaks down in the system K', and so too we are unable to introduce a time corresponding to physical requirements in K', indicated by clocks at rest relatively to K.

*Einstein used the effects from special relativity only in order to convince the reader of these two conclusions.* Einstein was going to present a new theory, from which he would explain effects such as those exemplified in the disk thought experiment in whole new manner.

A few months later, after writing the 1916 review article, the disk thought experiment reappeared in Einstein's popular book, *Über die spezielle und die allgemeine Relativitätstheorie (Gemeinverständlich)* [On the Special and General Theories of Relativity (General Course)]. In section §23 "Behavior of Clocks and Rods on a Rotating Reference-Body" Einstein presented the Disk story. Einstein said he starts from a special case upon which he had already frequently relied.[59]

He considered a reference body K lying in a region in space-time in which no gravitational field exists. K is a Galilean reference body, and the results of special relativity are valid with respect to K. Now consider in the same region another reference body, K', which is uniformly rotating relative to K. Einstein assumed that K' is ebenen Kreisscheibe (plane circular disk), which rotates uniformly about its center point.[60] And the disk thought experiment continues in quite the same manner as in the 1916 review article.[61] However, in the 1916 review article Einstein did not explicitly refer to an ebenen Kreisscheibe; he spoke of a "circle", even though it was implicit to the discussion. And in the 1916 book Einstein framed the discussion of the disk thought experiment in terms of a plane circular disk. Einstein thus tried to identify the

Disk with a real object, and so the empirical foundations of his theory would be more secure. [62]

## 2.6 The Point Coincidence Argument

The main conclusion derived from the disk thought experiment was that, in the general theory of relativity the method of laying coordinates in the space-time continuum in a definite manner breaks down, and one cannot adapt coordinate systems to the four-dimensional space. Einstein thus arrived at the conclusion that there was no way of arriving at a simple formulation of the laws of nature.

This conclusion brought Einstein to a formulation of a principle of general covariance: If we cannot be dependent on the above space and time measurements, then we must regard all imaginable systems of coordinates, on principle, as equally suitable for the description of nature:[63]

"*The general laws of nature have to be expressed by equations which are valid for all coordinate systems, i.e., are covariant with respect to any substitutions (generally covariant)*".

This postulate did not yet include field equations and it was a coordinate-dependent version of the principle of general covariance.[64]

Einstein explained that, "It is clear that a physics that satisfies this postulate will do justice to the general postulate of relativity". And then Einstein presented the argument that supported this principle of general covariance:"That this natural requirement of general covariance, which takes away from space and time the last remnant of physical objectivity, can be seen from the following considerations. All our space-time verifications invariably amount to a determination of space-time coincidences".[65] Stachel showed in 1980-1989 that Einstein presented this Point Coincidence Argument instead of the Hole Argument in section §12 of the 1914 review article.[66] The context in which Einstein presented the point coincidence argument was his rejection of his Hole Argument and his commitment and acceptance of general covariance.[67]

There are actually two versions of the Point Coincidence Argument, which have been called "the private" and "the public" one.[68] The private version was formulated in letters. Recall Ehrenfest's query in his letter to Einstein about the 1914 Hole. Einstein in reply explained to Ehrenfest and also to his best friend Michele Besso that everything was correct with the Hole Argument up to the final conclusion afterwhich,[69]

"The hole argument is replaced by the following consideration. Nothing is physically *real* but the totality of space-time point coincidences. If, for example, all physical events were to be built up from the motions of material points alone, then the meetings of these points, i.e., the points of intersection of the world lines, would be the only real things, i.e., observable in principle. These points of intersection naturally are preserved during all [coordinate] transformations (and no new ones occur) if only certain uniqueness conditions are observed. It is therefore most natural to demand of the laws that they determine no *more* than the totality of space-time coincidences. From what has been said, this is already attained through the use of generally covariant equations".

Einstein explained his public version of the Point Coincidence Argument in print in the 1916 paper in the following way.

First, suppose that all events consisted only of the motion of material points. Therefore, Einstein's theory only determines the meetings of two or more of these points, the "space-time coincidences". However, here comes the problematic definition: The results of our measurements are verifications of these meetings of the material points with the material points of our measuring instruments. That is, our measurements consist of "coincidences between the hands of a clock and points on the clock dial, and observed point-events happening at the same time".[70]

This definition is problematic, because it should take into account the definition of simultaneity. Recall the definition from the first 1905 special relativity paper, "If for instance, I say, 'That train arrives here at 7 o'clock', I mean something like this: 'This pointing of the small hand of my watch to 7 and the arrival of the train are simultaneous events".[71] However, in special relativity coordinates of space and time have direct physical meaning. If a point has the coordinate $x_4 = t$, this means that "at rest with respect to the coordinate system, a standard unit clock, which is (practically) coincident with the point event, will have measured off $x_4 = t$ periods at the occurrence of the point-event".[72] On the basis of these assumptions, and generally the assumptions of special relativity, one can define simultaneity. But what about general relativity? At a footnote Einstein wrote that we assume the possibility of verifying simultaneity for events immediately adjacent in space (or for coincidences in space-time), without giving a definition of this fundamental concept.[73]

Subsequently Einstein presented the Point-Coincidence Argument,

We associate to the world four space-time variables $x_1$, $x_2$, $x_3$, $x_4$. For every point-event there is a corresponding system of values of the variables $x_1 \ldots x_4$. We associate the above variables, coordinates, to space-time coincidences of point-events, Einstein then wrote, "Two coincident point-events correspond to the same system of values of the variables $x_1 \ldots x_4$, i. e., the coincidence is characterized by the identity of the coordinates. If, instead of the variables $x_1 \ldots x_4$, we introduce functions of them, $x'_1$, $x'_2$, $x'_3$, $x'_4$, as a new coordinate system, so that the system of values corresponds to one another unambiguously, then the equality of all four coordinates in the new system will also serve as an expression for the space-time coincidence of the two point-events. Since all our physical experience can be ultimately reduced to such point coincidences, there is no immediate reason for preferring certain systems of coordinates to others, i.e., we arrive at the requirement of general covariance".[74]

In the Point-Coincidence Argument Einstein represented space-time coincidences by variables, and then he presented a new coordinate system: the functions of the variables. The system of these values corresponds to one another without ambiguity. And there is no immediate reason for preferring certain systems of coordinates to others. Einstein was presenting his argument in terms of a system of particles, rather than fields, the model being any set of particle world lines, without any requirement that they satisfy equations of motion; Einstein did not mention any dynamical equations or even fields; and the objects did not necessarily obey any field equations.[75]

How did Einstein avoid the Hole Argument? In the 1914 Hole Argument the physical events are completely determined if the quantities $g'_{\mu\nu}$ are given as functions of the $x'_\nu$ with respect to the coordinate system K' used for the description, symbolically denoted by G'(x'). Then Einstein formed another function G'(x), which also describes a gravitational field with respect to K. Considering generally covariant field equations, then with respect to K there exist two different solutions G(x) and G'(x), which are different from one another, but at the boundary of the hole both solutions coincide.

In the Hole Argument G(x) and G'(x) represented two different gravitational fields with respect to K; but according to the Point Coincidence Argument, G(x) and G'(x) should represent the same gravitational field: Recall that in the Point Coincidence Argument all events consisted only of the motion of material points, and we are dealing only with the coincidences of the space-time points. Suppose we take the first solution – gravitational field – G(x) and think of it as material points moving in space-time. Then according to Einstein's above definition we are dealing only with the meeting of the two material points. Let us designate this meeting point by ($x_1$, $x_2$, $x_3$, $x_4$). Now we shall take the second solution G'(x) and think of it as well as material points moving in space-time. This latter according to Einstein's Point Coincidence Argument would be also reduced to the meeting of two material points at ($x'_1$, $x'_2$, $x'_3$, $x'_4$). The two points are indistinguishable, because there is no reason to preferring the first to the latter.

At this point of the paper, after presenting the Point Coincidence Argument, Einstein very likely intended to include another section under the title "The Fundamental property of Mass", and then he regretted. In the manuscript of the 1916 paper he wrote the following title on page 7,

"§4 Die fundamentale Mass-Eigenschaft" and crossed out this sentence,[76]

One can speculate about this title: In section §4 Einstein could have thought of formulating an additional principle, an initial version of Mach's principle. Or else he might have had initial cosmological thoughts, and he could have decided to postpone them to later writings.

Instead of the above title, Einstein immediately wrote another title for section §4:

"§4 The Relationship of Four Coordinates to Special and Temporal Measurements. Analytical Expressions for the Gravitational Field".[77]

This section was completely unrelated to Part A. It is thus evident that it came instead of another section, "The Fundamental property of Mass" section. The above section §4 that appears in the printed version in Part A is more naturally related to the next part, Part B.

## 2.7 The General Theory of Relativity of 1916: Natural not Simple

Einstein started section §4 of his theory with an excuse or an apology to the reader who began reading the mathematical part of the theory after he had read the heuristic part, "It seems to me unimportant in this treatise to represent the general theory of relativity as a system that is as simple and logical as possible and with minimum axioms". This was opposed to Einstein's coordinate-dependent *heuristic works* on gravitation from 1907-1912, characterized by the tacit assumption that coordinates and time measurements had direct physical meaning; Einstein was there searching for the most simple and logical system as possible and with minimum assumptions. As opposed to these works, Einstein's main object in the 1916 general theory of relativity was to develop a theory that the chosen path entered to it was *psychologically the natural one*, and its underlying assumptions would appear to have been *secured experimentally.*[78]

Indeed at first the general public perceived the General Theory of Relativity to be contra every-day common-sense. Einstein wrote in the forward to Philip Frank's biography,[79]

"For me it was always difficult to understand why in practical life the theory of relativity used so distant concepts and issues, and for so long caused the broadest sections of the working population so many problems, and indeed caused such a suffering response."

["Mir selbst war es stets unverständlich, warum die Relativitätstheorie mit ihren dem praktischen Leben so entfernten Begriffen und Problemstellungen in den breitesten Schichten der Bevölkerung für eine lange Zeit eine so lebhafte, ja leidenschaftliche Resonanz gefunden hat"].

Einstein said that he chose the natural path. A natural path was to start with special realtivity as the limiting case, the *flat Minkowski metric*,[80]

"In this sense this condition is introduced: For infinitely small four dimensional regions the theory of relativity in the restricted sense is applicable at an appropriate choice of the coordinates".

However, Einstein followed his 1914 review article and chose an infinitely small or local coordinate system in such a way, with almost no acceleration, so that *no gravitational field occurs*. In section §4 of the 1916 paper he gave the same equation that he had presented in his 1914 review article in section §2, "The Gravitational Field", equation (2b). But he added a new equation (1) and an important explanation:[81]

Suppose $X_\nu$ ($\nu$ = 1, 2, 3,4) are coordinates of space and time used in the infinitely small four dimensional local regions, and they have direct physical meaning for this infinitely small area (according to special relativity). They are measured with a rigid rod (and a clock). Einstein then wrote the expression,

$$(1)\ ds^2 = -\,dX_1^2 - dX_2^2 - dX_3^2 + dX_4^2.$$

Hence for local systems,

$$ds^2 = -\sum dX_\nu^2 \ (\nu = 1,2,3,4).$$

In the 1914 paper Einstein did not yet emphasize that $X_\nu$ have direct physical meaning in the sense of the special theory of relativity. Einstein arrived at an understanding that in the general theory of relativity space and time coordinates have no direct physical meaning only after he abandoned the Hole Argument.

For the whole region Einstein wrote,

$$(3)\ ds^2 = \sum_{\sigma\tau} g_{\sigma\tau} dx_\sigma dx_\tau.$$

Where, $g_{\sigma\tau}$ are functions of the $x_\sigma$.[82]

One obtains the "Usual Theory of Relativity" (special relativity) for local systems when the $g_{\sigma\tau}$ are equal to the following constant values:[83]

(4) $g_{\sigma\tau} = \text{diag}(-1, -1, -1, +1)$.

This is *Minkowski flat metric*. A free material point then moves with respect to this local system uniformly in a straight line.[84]

However, once we introduce new space-time coordinates $x_\sigma$ ($\sigma$=1, 2, 3, 4), by substitution, then the $g_{\sigma\tau}$ in this system *are not constants any more*, *but functions of space-time*. From the physical point of view, the quantities $g_{\sigma\tau}$ describe the gravitational field with respect to the chosen reference system.[85]

In general relativity, "Gravitation plays, therefore, an exceptional role according to the general theory of relativity, with regard to other forces, especially the electromagnetic forces, since the 10 functions $g_{\mu\nu}$ representing the gravitational field also determine the metrical properties of the four-dimensional space measured".[86]

**2.8 Mathematical Aids and the Summation Convention**

Einstein now arrived at section B "Mathematical Aids to the Formulation of Generally Covariant Equations".[87] This section exchanges section B of the 1914 review article "From the Theory of Covariants".[88] However, in 1916, part B was meant to assist to the generally covariant theory and thus included some of the insights from the November 1915 papers. The purpose of part B is: "By examining the laws of the formulation of tensors, we obtain the means of establishing generally covariant laws".[89]

A novel tool that Einstein introduced in 1916 was the "Einstein summation convention". In the printed version of the 1916 paper the summation convention is highlighted: "*Bemarkungzur vereinfachungder Schreibweise der Ausdrücke*."[90] In the manuscript of the 1916 paper, Einstein did not highlight this sentence (did not draw any line beneath this sentence).[91] Hence it is reasonable that after Einstein had submitted the paper to the *Annalen*, he thought that his idea about the summation convention was very important, and thus before the paper was printed he corrected this and asked to highlight this sentence.

Einstein wrote: "It is therefore possible, without loss of clearness, to omit the sign of summation. For this purpose we introduce the rule: If an index occurs twice in one term of an expression, it is always to be summed unless the contrary is expressly stated".[92] Thus the index $\sigma$ is free and it can get any value from 1 to n, because we do not sum upon it with $\Sigma$, and the index $\nu$ is dummy and thus is summed over 1 to n.

All the equations of the 1914 review article and the November 1915 papers differ from those of the 1916 review article because of the summation convention. Einstein took the equations from his 1914 article and from his November 1915 papers and omitted the summation sign. It was not only an "aesthetic" change, but it also simplified calculations. Although in many cases – as we shall see – Einstein simply transferred equations from the above sources to his 1916 paper – he actually presented quite a new mathematical formulation.

Einstein wrote equations (5a) and (8) from his 1914 review article.[93] Here, the transformation law for contravariant tensors of the second rank (using the summation convention):[94]

$$(9)\ A^{\sigma\tau\prime} = \frac{\partial x'_{\sigma}}{\partial x_{\mu}} \frac{\partial x'_{\tau}}{\partial x_{\nu}} A^{\mu\nu},$$

and the transformation law for covariant tensors of the second rank:[95]

$$(11)\ A_{\sigma\tau}{}' = \frac{\partial x_{\mu}}{\partial x'_{\sigma}} \frac{\partial x_{\nu}}{\partial x'_{\tau}} A_{\mu\nu}.$$

Now Einstein added something he had not included in his 1914 review article. He wrote the equation for the transformation law for mixed tensors, which are covariant with respect to the index μ, and contravariant with respect to the index ν,[96]

$$(13)\ A_{\sigma}^{\tau}{}' = \frac{\partial x'_{\tau}}{\partial x_{\beta}} \frac{\partial x_{\alpha}}{\partial x'_{\sigma}} A_{\alpha}^{\beta}.$$

In the 1914 review article, this equation was written in the form of equation (9).[97]

Einstein then presented *symmetrical tensors*, which he did not present in his 1914 review article. He could now say he wanted to prove that symmetry is a property which is independent of the reference system. Using (9) and the symmetry condition for tensors, $A^{\mu\nu} = A^{\nu\mu}$, Einstein managed to prove that, $A^{\sigma\tau\prime} = A^{\tau\sigma\prime}$.[98] He ended the section with anti-symmetric tensors,[99] which he had presented in his 1914 review article.[100]

Einstein arrived at the next section §7 "Multiplication of tensors".[101] This section also appeared in the 1914 paper after presenting the anti-symmetric tensors. Einstein rewrote this mathematical section of his 1914 paper using the summation convention and in light of his November 1915 papers. This is also seen in section §8; in this section Einstein discussed "Some Aspects of the Fundamental Tensor $g_{\mu\nu}$".[102] This section was a correction of the equivalent section §6 of the 1914 paper ("On Some Relations Concerning the Fundamental Tensor $g_{\mu\nu}$") that came right after the 1914 section §5 "Multiplication of Tensors".[103]

In section §8 Einstein wrote some relations concerning the (metric) fundamental tensor $g_{\mu\nu}$.[104] Einstein wrote again the 1914 equations (10) ("a known property of determinants"):[105]

$$\left|g_{\mu\alpha}g^{\alpha\nu}\right| = \left|\delta_\mu^\nu\right| = 1,$$

the unsigned equations that followed (10),

$$\left|g_{\mu\alpha}g^{\alpha\nu}\right| = \left|g_{\mu\nu}\right|\left|g^{\mu\nu}\right| = \left|\delta_\mu^\nu\right| = 1,$$

and the multiplication theorem of determinants, which led to the 1914 equation (11) for the fundamental tensor, [106]

$$(17)\left|g_{\mu\nu}\right|\left|g^{\mu\nu}\right| = 1.$$

Einstein rewrote the equations from page 1041 of the 1914 paper; but instead of the 1914 equation [(14) $d\tau_0^* = |\alpha_{\sigma\mu}|d\tau$] he put the equation from his November 4, 1915 paper, [107]

$$[(17a)]\ d\tau' = \left|\frac{\partial x'_\sigma}{\partial x_\mu}\right| d\tau.$$

And,

$$d\tau' = \int dx_1\, dx_2 dx_3 dx_4.$$

Einstein then corrected his 1914 procedure, [108]

$$\sqrt{g'} = \left|\frac{\partial x_\mu}{\partial x'_\sigma}\right| \sqrt{g}.$$

From these we obtain,

$$(18)\ \sqrt{g'}d\tau' = \sqrt{g}d\tau.$$

This corrected the 1914 equation (17). This latter invariant is equal to the magnitude of the four-dimensional volume element in the local reference system, as measured with rigid rods and clocks in the sense of the special theory of relativity.

If $d\tau_0$ is the "natural" volume element *in the local reference system where special relativity* applies, then,[109]

(18a) $d\tau_0 = \sqrt{-g}d\tau.$

This is equation (17a) from the 1914 paper.

If $\sqrt{-g} = 1$, then $d\tau_0 = d\tau$.

Let us come back to Einstein's definition from the beginning of section §4, which is actually the equivalence principle applicable to local systems: Experiments in a sufficiently small free falling system, over a sufficiently short time interval, give results that are indistinguishable from those of the same experiments done in an inertial frame in which special relativity applies.

Consider the following *Gedanken*-experiment. Imagine two systems. One system is Einstein's imaginary man falling from the roof in a gravitational field, and in the other system there is a man at rest in a gravitational field. Let us consider the local inertial system of special relativity: Imagine the man at the moment he starts to fall from the roof. At this infinitesimally initial moment, he is still at rest. Both men at both systems are therefore at rest at this very moment. The worldlines of these men are comprised of the time intervals according to (18): $\sqrt{g_{44}{}'}d\tau' = \sqrt{g_{44}}d\tau$. In the local inertial system all the diagonal components of the metric tensor are constants and according to (4) they are equal to 1, and the off diagonal components are zero. And thus $g_{44} = 1$, and we arrive at an equation (18a). This expression is very similar to the time transformation from the special theory of relativity:

$$\sqrt{1 - \mathrm{v}^2/c^2}\, d\tau = d\tau_0.$$

Einstein wrote in his Prague 1911 paper, "If we measure time in $S_1$ with a clock *U*, *we must measure the time in* $S_2$ *with a clock that goes* $1 + \Phi/c^2$ *slower than the clock U if you compare it with the clock U in the same place*".[110] Thus,

$(1 + \Phi/c^2)t_1 = t_2$.

In the limit of weak gravitational fields, Einstein expected that $g_{44}$ would tend to the above factor. Thus:

$$\sqrt{g_{44}} = \sqrt{1 + \frac{\Phi}{c^2}}.$$

As Einstein showed in his November 4 paper, it is possible to achieve a simplification of the laws of nature if in place of (18) we have simply dτ' = dτ, from which in view of (17a), it follows that,

$$(19)\ \left|\frac{\partial x_\sigma'}{\partial x_\mu}\right| = 1.$$

Thus with this choice of coordinates, only substitutions for which the determinant is unity are permissible.

Einstein then explained why this choice of November 4 does not mean a "partial abandonment of the general postulate of relativity". The reason is that we do not ask what are the laws of nature which are covariant for transformations of determinant 1? We rather first ask: "What are the generally covariant laws of nature?" [111] Only after formulating these, we then simplify their expression by a special choice of a reference system (or come back to some form of Einstein's Angepaßte Koordinatensysteme). In the 1914 paper Einstein did not possess any generally covariant field equations, and he could only prove covariance with respect to the adapted coordinate system. This was indeed "partial abandonment of the general postulate of relativity". Hence, following the stages from the beginning of November 1915, Einstein adopted the determinant in equation $\sqrt{-g} = 1$ as a postulate. Then in the November 11 paper he adopted it as a coordinate condition, and in the 1916 review article he expressed his field equations with respect to the special reference system $\sqrt{-g} = 1$.[112]

In the 1914 review article after writing the equivalent equation to equation (18a), Einstein presented the Ricci tensor. It was a mathematical presentation with no physical significance. In the 1916 review article this tensor was already central to Einstein's theory, and thus the whole presentation of this tensor was of course different and discussed in a separated section. Before presenting the Ricci tensor, Einstein discussed the geodesic line.

In section §9 Einstein derived "The equation of the Geodesic Line. The Motion of a Particle".[113] This section was equivalent to section §7 from the 1914 paper "The Geodesic Line and Equations of Motions of Points".[114] In fact the title of both sections is quite similar. And in addition, Einstein copied the equations from section §7 of the 1914 paper into section §9 of the 1916 paper, but in the latter he used the summation convention and indeed, as said already, this is the difference between the equations of the two sections.

**2.9 Geodesic Equation**

Further below, in Part C, in section §13 "Equations of motion of a material point in the gravitational field", Einstein wrote the geodesic equations from his November 4, 1915 paper!

Einstein wrote the 1914 equation (23a), the differential equations that define the geodesic line,[115]

$$(20d)\quad g_{\alpha\sigma}\frac{d^2x_\alpha}{ds^2}+\begin{bmatrix}\mu\nu\\ \sigma\end{bmatrix}\frac{dx_\mu}{ds}\frac{dx_\nu}{ds}=0,$$

And the 1914 equations (24), the Christoffel symbols of the first kind,

$$(21)\quad \begin{bmatrix}\mu\nu\\ \sigma\end{bmatrix}=\frac{1}{2}\left(\frac{\partial g_{\mu\sigma}}{\partial x_\nu}+\frac{\partial g_{\nu\sigma}}{\partial x_\mu}-\frac{\partial g_{\mu\nu}}{\partial x_\sigma}\right).$$

By contraction, multiplying (20d) by $g^{\sigma\tau}$ (outer multiplication with respect to τ, inner with respect to σ), $g_{\alpha\sigma}g^{\sigma\tau}=\delta_\alpha^\tau=1$, we obtain the equations of the geodesic line – the 1914 equations (23b) – in the following form,[116]

$$(22)\quad \frac{d^2x_\tau}{ds^2}+\begin{Bmatrix}\mu\nu\\ \tau\end{Bmatrix}\frac{dx_\mu}{ds}\frac{dx_\nu}{ds}=0,$$

where the Christoffel symbols of the second kind the 1914 equations (24b) are,

$$(23)\quad \begin{Bmatrix}\mu\nu\\ \tau\end{Bmatrix}=g^{\tau\alpha}\begin{bmatrix}\mu\nu\\ \alpha\end{bmatrix}.$$

**2.10 Tensors Again**

After obtaining the geodesic equation Einstein reproduced the section that came after this one in his 1914 paper. In his 1914 paper the next section was "§8 Forming Tensors by Differentiation", and in the 1916 paper it was "§10 The Formation of Tensors by Differentiation".[117] Therefore most of the equations in both sections are equivalent, except the summation convention. The important equations from these sections are the following.

In 1914 Einstein wanted to derive the laws by which new tensors can be formed from known tensors by differentiation.[118]

Suppose that S is a given curve in the continuum. It has a length of arc s from a point P on S. The quantity φ is (an invariant) a scalar, and thus the quantities dφ/ds, $d^2\varphi/ds^2$, etc are also scalars, quantities that are independent of the coordinate system. [119]

$$\frac{d\varphi}{ds} = \frac{\partial\varphi}{\partial x_\mu}\frac{dx_\mu}{ds},$$

and the invariant (or scalar),

$$\psi = \frac{\partial\varphi}{\partial x_\mu}\frac{dx_\mu}{ds}.$$

Thus, $\psi = A_\mu \frac{dx_\mu}{ds} = \frac{d\varphi}{ds}, \; A_\mu = \frac{\partial\varphi}{\partial s}\frac{ds}{dx_\mu} = \frac{\partial\varphi}{\partial x_\mu}.$

And Einstein obtained the following covariant vector,

$$(24) \; A_\mu = \frac{\partial\varphi}{\partial x_\mu}.$$

Einstein said in 1914 that we now assume that the line S is a geodesic. This choice is independent of the adopted reference system.

If we differentiate ψ with respect to S,

$$\frac{d^2\varphi}{ds^2} = \frac{d\psi}{ds}$$

then we would not obtain a differential which is a tensor. However if the curve along which one performs this differentiation is a geodesic, then the differential is a tensor, a covariant tensor of the second rank. We thus use equation (22) – the differential equation of the geodesic – and the value for ψ, and insert them into the above equation and obtain, [120]

$$\frac{d^2\varphi}{ds^2} = \left\{\frac{\partial^2\varphi}{\partial x_\mu \partial x_\nu} - \begin{Bmatrix}\mu\nu \\ \tau\end{Bmatrix}\frac{\partial\varphi}{\partial x_\tau}\right\}\frac{dx_\mu}{ds}\frac{dx_\nu}{ds}.$$

Einstein then focused on the quantities in the brackets of the above formula and wrote the covariant tensor of the second rank,

$$(25)\ A_{\mu\nu} = \frac{\partial^2 \varphi}{\partial x_\mu \partial x_\nu} - \begin{Bmatrix} \mu\nu \\ \tau \end{Bmatrix} \frac{\partial \varphi}{\partial x_\tau}.$$

Thus from the covariant tensor of the first rank (24) we can by differentiation form a covariant tensor of the second rank,

$$\frac{\partial A_{\mu\nu}}{\partial x_\mu} = \frac{\partial^2 \varphi}{\partial x_\mu \partial x_\nu}$$

And thus,[121]

$$(26)\ A_{\mu\nu} = \frac{\partial A_\mu}{\partial x_\mu} - \begin{Bmatrix} \mu\nu \\ \tau \end{Bmatrix} A_\tau.$$

This tensor is the extension – covariant derivative – of the tensor $A_\mu$.
And the extension of the tensor $A_{\mu\nu}$ is,[122]

$$(27)\ A_{\mu\nu\sigma} = \frac{\partial A_{\mu\nu}}{\partial x_\sigma} - \begin{Bmatrix} \sigma\mu \\ \tau \end{Bmatrix} A_{\tau\nu} - \begin{Bmatrix} \sigma\nu \\ \tau \end{Bmatrix} A_{\mu\tau}.$$

In the manuscript of the 1916 paper Einstein first wrote this equation in the following way,

$$A_{\mu\nu\sigma} = \frac{\partial A_{\mu\nu}}{\partial x_\sigma} - \begin{Bmatrix} \sigma\mu \\ \tau \end{Bmatrix} \mathbf{A}^{\nu}_{\tau\mu} - \begin{Bmatrix} \sigma\nu \\ \tau \end{Bmatrix} A_{\mu\tau},$$

and then he erased the term $A^{\nu}_{\tau\mu}$ in the second term on the right hand side and wrote $A_{\tau\nu}$ instead of it as it appears in the above equation (27).[123] Einstein was then thinking of extension of mixed tensors, equation (39) below, and decided to postpone it to the next section §11.

Generally the formula for the extension of any covariant tensor of rank *l* in order to obtain a tensor of rank (*l* + 1) is equation (29) from the 1914 paper, and equation (30) of that paper for contravariant tensors.[124]

Section §11 of the 1916 paper discussed "some cases of Special importance", and it included some of the formulas from section §8 of the 1914 paper.[125]

Einstein first wrote the rule for the differentiation of determinants for the fundamental tensor,[126]

$$(28)\ dg = g^{\mu\nu} g dg^{\mu\nu} = -g_{\mu\nu} g dg^{\mu\nu}.$$

Or,

$$\frac{dg}{dx_\sigma} = g^{\mu\nu} g \frac{dg^{\mu\nu}}{dx_\sigma}.$$

Then from (28) it follows,

$$(29) \frac{1}{\sqrt{-g}} \frac{\partial\sqrt{-g}}{\partial x_\sigma} = \frac{1}{2} g^{\mu\nu} \frac{\partial g_{\mu\nu}}{\partial x_\sigma} = -\frac{1}{2} g_{\mu\nu} \frac{\partial g^{\mu\nu}}{\partial x_\sigma}.$$

In the manuscript of the 1916 paper Einstein wrote this equation with another 1/2 and crossed out it,[127]

$$\frac{1}{\sqrt{-g}} \frac{\partial\sqrt{-g}}{\partial x_\sigma} = \mathbf{\frac{1}{2}} \frac{1}{2} g^{\mu\nu} \frac{\partial g_{\mu\nu}}{\partial x_\sigma} = -\frac{1}{2} g_{\mu\nu} \frac{\partial g^{\mu\nu}}{\partial x_\sigma}.$$

From, $g_{\mu\sigma} g^{\nu\sigma} = \delta_\mu^\nu$ it follows on differentiation that,[128]

$$(30)\quad g_{\mu\sigma} dg^{\nu\sigma} = -g^{\nu\sigma} dg_{\nu\sigma}, \text{ or,}$$

$$g_{\mu\sigma} \frac{\partial g^{\nu\sigma}}{\partial x_\lambda} = -g^{\nu\sigma} \frac{\partial g_{\mu\sigma}}{\partial x_\lambda}.$$

Einstein then obtained additional relations he would use later, especially the first of these relations and the equations derived from it using the Christoffel symbols (21) and (23),[129]

$$(31)\quad dg^{\mu\nu} = -g^{\mu\alpha} g^{\nu\beta} dg_{\alpha\beta},$$

$$\frac{\partial g_{\mu\nu}}{\partial x_\sigma} = -g^{\mu\alpha} g^{\nu\beta} \frac{\partial g_{\alpha\beta}}{\partial x_\sigma}.$$

Or,

$$(32)\ dg_{\mu\nu} = -g_{\mu\alpha} g_{\nu\beta} dg^{\alpha\beta},$$

$$\frac{\partial g_{\mu\nu}}{\partial x_\sigma} = -g_{\mu\alpha} g_{\nu\beta} \frac{\partial g^{\alpha\beta}}{\partial x_\sigma}.$$

Using the Christoffel symbols of the first kind (21), the second formula of (31) transform in the following way,

$$(33)\ \frac{\partial g_{\alpha\beta}}{\partial x_\sigma} = \begin{bmatrix} \alpha\sigma \\ \beta \end{bmatrix} + \begin{bmatrix} \beta\sigma \\ \alpha \end{bmatrix}.$$

Inserting this in the second formula of (31) in view of the Christoffel symbols of the second kind (23), one obtains,

$$(34)\ \frac{\partial g^{\mu\nu}}{\partial x_\sigma} = -\left( g^{\mu\tau} \begin{Bmatrix} \tau\sigma \\ \nu \end{Bmatrix} - g^{\nu\tau} \begin{Bmatrix} \tau\sigma \\ \mu \end{Bmatrix} \right).$$

Substituting the right-hand side of this equation in (29) leads to,

$$(29a)\ \frac{1}{\sqrt{-g}} \frac{\partial \sqrt{-g}}{\partial x_\sigma} = \begin{Bmatrix} \mu\sigma \\ \mu \end{Bmatrix}.$$

The above highlighted equations appeared in the manuscript on a small piece of paper that was glued to the bottom of sheet (page) number 23. Einstein presumably did not intend to include these equations, or else he forgot about them when he wrote on clean and nice sheets his paper. And then all of a sudden he remembered that he forgot these equations, and then included them; and thus they are related to page 23. The more reasonable hypothesis is that Einstein did not intend to include these three equations in the text, and only later did he decide to include them. The reason for this is that later on in the paper Einstein uses these equations. Hence, when Einstein realized, that he would need these equations for the derivations of his theory, he took a small piece of paper, wrote these three equations and glued it to page 23. [130]

In addition, he wrote the number (23a) above equation (33). Einstein thus signified to the editor of the *Annalen* that the small piece of paper and the three equations are page 23a and it is related to page (23). [131]

These equations had already appeared in the 1914 paper. Equation (28) from 1916 was written in 1914 as equation (34). [132] Equations (29) and (29a) are combined in the 1914 paper into equation (33).[133] And equations (31) and (32) were written in 1914 as equations (35) and (36).[134]

Einstein next returned to the covariant derivative (26). Using (31) and (29) he obtained the scalar equation (37) of his 1914 paper,[135] the divergence of the contravariant vector $A^\nu$.[136]

$$(35)\ \Phi = \frac{1}{\sqrt{-g}}\frac{\partial}{\partial x_\nu}\left(\sqrt{-g}\,A^\nu\right).$$

In his November 4 paper, Einstein assumed the determinant 1 condition, and he thus replaced this equation with a simpler equation by omitting $\sqrt{-g}$.[137]

Einstein derived two extra important relations. He first rewrote (27) in a new form and used (34) and (29a). He obtained,[138]

$$(38)\ A^{\alpha\beta}_{\sigma} = \frac{\partial A^{\alpha\beta}}{\partial x_\sigma} + \begin{Bmatrix}\sigma\chi\\ \alpha\end{Bmatrix}A^{\chi\beta} + \begin{Bmatrix}\sigma\chi\\ \beta\end{Bmatrix}A^{\alpha\chi}.$$

This is the expression for the extension of a contravariant tensor of the second rank. And this is the extension of a mixed tensor,

$$(39)\ A^{\alpha}_{\mu\sigma} = \frac{\partial A^{\alpha}_{\mu}}{\partial x_\sigma} - \begin{Bmatrix}\sigma\mu\\ \tau\end{Bmatrix}A^{\alpha}_{\tau} + \begin{Bmatrix}\sigma\tau\\ \alpha\end{Bmatrix}A^{\tau}_{\mu}.$$

From (38) and (29a) Einstein obtained the divergence of a contravariant six vector,[139]

$$(40)\ A^{\alpha} = \frac{1}{\sqrt{-g}}\frac{\partial\left(\sqrt{-g}A^{\alpha\beta}\right)}{\partial x_\beta}.$$

From (39) and (29a) he obtained the divergence of a mixed tensor of the second rank,[140]

$$(41)\ \sqrt{-g}A_\mu = \frac{\partial\left(\sqrt{-g}A^{\sigma}_{\mu}\right)}{\partial x_\sigma} - \begin{Bmatrix}\sigma\mu\\ \tau\end{Bmatrix}\sqrt{-g}A^{\sigma}_{\tau}.$$

**2.11 The Einstein tensor**

In section §8 of the 1914 paper (on page 1053) Einstein presented "*The Riemann-Christoffel Tensor*". [141] Recall that he did not use this tensor in his "Entwurf" field equations of 1914; hence the exposition in 1914 was brief. Subsequently Einstein gave another exposition, the V-tensors (tensor densities) [142].

In the 1916 paper the Riemann-Christofell Tensor occupied a whole section, section §12. [143] Einstein started this section with the two equations he had written on page 1053 of his 1914 paper, while the second of these equations was the Riemann-Christoffel Tensor.[144] Then he explained the mathematical importance of this tensor, an explanation which was absent in the 1914 paper. Finally, the discussion in section §12 ended with the central equations appearing in all of the November 1915 papers, equations (13), (13a), and (13b). These three equations were written now with Einstein's summation convention rule. Einstein ended the section with the important note about the choice of coordinates such that $\sqrt{-g} = 1$.

In 1914 Einstein used equation (29) from this paper, the equation that extends a tensor of rank *l* to a tensor of rank *l*+1: Einstein formed the tensor $A_{\mu\nu}$ of the second rank from the covariant tensor of the first rank (covariant four vector) $A_\mu$. Equation (26) of the 1916 paper also represents an extension of a covariant four vector $A_\mu$ to a tensor $A_{\mu\nu}$ of the second rank. Einstein then extended (26) to the tensor of rank three $A_{\mu\sigma\nu}$ in (27). In (27) he placed the extension of the four vector $A_\mu$ (26). Then he obtained the following tensor of the third rank,[145]

$$A_{\mu\sigma\tau} = \frac{\partial^2 A_\mu}{\partial x_\sigma \partial x_\tau} - \begin{Bmatrix} \mu\sigma \\ \rho \end{Bmatrix} \frac{\partial A_\rho}{\partial x_\tau} - \begin{Bmatrix} \mu\tau \\ \rho \end{Bmatrix} \frac{\partial A_\rho}{\partial x_\sigma} - \begin{Bmatrix} \sigma\tau \\ \rho \end{Bmatrix} \frac{\partial A_\mu}{\partial x_\rho}$$
$$+ \left[ -\frac{\partial}{\partial x_\tau} \begin{Bmatrix} \mu\sigma \\ \rho \end{Bmatrix} + \begin{Bmatrix} \mu\tau \\ \alpha \end{Bmatrix} \begin{Bmatrix} \alpha\sigma \\ \rho \end{Bmatrix} + \begin{Bmatrix} \sigma\tau \\ \alpha \end{Bmatrix} \begin{Bmatrix} \alpha\mu \\ \rho \end{Bmatrix} \right] A_\rho .$$

From this expression it is seen that we can also form the following symmetrical covariant tensor of rank three, $A_{\mu\sigma\tau} - A_{\mu\tau\sigma}$. Then some of the terms in the expression of $A_{\mu\sigma\tau}$ cancel those in the expression of $A_{\mu\tau\sigma}$: The first term, $\frac{\partial^2 A_\mu}{\partial x_\sigma \partial x_\tau}$, the fourth term, $\begin{Bmatrix} \sigma\tau \\ \rho \end{Bmatrix} \frac{\partial A_\mu}{\partial x_\rho}$, and some of the last terms in the square brackets, because all these are symmetrical in the indices σ and τ. Thus Einstein wrote,

$$(42)\ A_{\mu\sigma\tau} - A_{\mu\tau\sigma} = B^\rho_{\mu\sigma\tau} A_\rho$$

$$(43)\ B^\rho_{\mu\sigma\tau} = -\frac{\partial}{\partial x_\tau} \begin{Bmatrix} \mu\sigma \\ \rho \end{Bmatrix} + \frac{\partial}{\partial x_\sigma} \begin{Bmatrix} \mu\tau \\ \rho \end{Bmatrix} - \begin{Bmatrix} \mu\sigma \\ \alpha \end{Bmatrix} \begin{Bmatrix} \alpha\tau \\ \rho \end{Bmatrix} + \begin{Bmatrix} \mu\tau \\ \alpha \end{Bmatrix} \begin{Bmatrix} \alpha\sigma \\ \rho \end{Bmatrix}$$

This is the Riemann-Christoffel Tensor, a mixed tensor of rank four. [146]

In 1914 Einstein presented the Riemann-Christoffel Tensor as a criterion that would allow him to decide whether a given line element is Euclidean. [147] In 1916 Einstein presented the same tensor for similar reasons. If there is a coordinate system in the continuum, with reference to which the $g_{\mu\nu}$ are constants, then $B^{\rho}_{\mu\sigma\tau}$ vanishes. The vanishing of the Riemann tensor $B^{\rho}_{\mu\sigma\tau}$ is a necessary condition that, by an appropriate choice of the reference systems, the $g_{\mu\nu}$ may be constants. "In our problem this corresponds to the case where, with a suitable choice of a coordinate system, the special theory of relativity is valid for finite areas".[148]

Einstein contracted (43) with respect to the indices τ and ρ, and obtained the Ricci tensor. Thus Einstein rewrote equation (13) from his 1915 papers with his summation convention, [149]

$$(44)\ B_{\mu\nu} \equiv G_{\mu\nu} = R_{\mu\nu} + S_{\mu\nu},$$

$$R_{\mu\nu} = -\frac{\partial}{\partial x_\alpha}\begin{Bmatrix}\mu\nu\\ \alpha\end{Bmatrix} + \begin{Bmatrix}\mu\alpha\\ \beta\end{Bmatrix}\begin{Bmatrix}\nu\beta\\ \alpha\end{Bmatrix}$$

$$S_{\mu\nu} = \frac{\partial \lg\sqrt{-g}}{\partial x_\mu \partial x_\nu} - \begin{Bmatrix}\mu\nu\\ \alpha\end{Bmatrix}\frac{\partial \lg\sqrt{-g}}{\partial x_\alpha}$$

In the manuscript of the 1916 paper Einstein wrote this last equation with an extra term that was crossed out (the highlighted term below):[150]

$$S_{\mu\nu} = \boldsymbol{\frac{\partial}{\partial x_\sigma}}\frac{\partial \lg\sqrt{-g}}{\partial x_\mu \partial x_\nu} - \begin{Bmatrix}\mu\nu\\ \alpha\end{Bmatrix}\frac{\partial \lg\sqrt{-g}}{\partial x_\alpha}$$

Therefore Einstein first intended to write this equation in the following form,

$$S_{\mu\nu} = \frac{\partial}{\partial x_\nu}\begin{Bmatrix}\mu\alpha\\ \alpha\end{Bmatrix} - \begin{Bmatrix}\mu\nu\\ \beta\end{Bmatrix}\begin{Bmatrix}\beta\alpha\\ \alpha\end{Bmatrix}$$

and according to equation (6) from the November 4 paper, [151]

$$\begin{Bmatrix} \mu\alpha \\ \alpha \end{Bmatrix} = \frac{1}{2} g^{\alpha\tau} \frac{\partial g_{\alpha\tau}}{\partial x_\mu} = \frac{\partial \log\sqrt{-g}}{\partial x_\mu}.$$

$G_{\mu\nu}$ (In 1916 Einstein wrote this tensor as $B_{\mu\nu}$ because of the notation that he gave to the Riemann tensor) is called today the *Einstein tensor*.

And the Einstein tensor with the components of the gravitational field,

$$G_{\mu\nu} \equiv B_{\mu\nu} = \frac{\partial \Gamma^{\alpha}_{\mu\nu}}{\partial x_\alpha} + \frac{\partial \Gamma^{\tau}_{\tau\mu}}{\partial x_\nu} + \Gamma^{\beta}_{\mu\alpha}\Gamma^{\alpha}_{\nu\beta} - \Gamma^{\alpha}_{\mu\nu}\Gamma^{\tau}_{\tau\alpha},$$

and splitting this tensor into two parts:

$$R_{\mu\nu} = -\frac{\partial \Gamma^{\alpha}_{\mu\nu}}{\partial x_\alpha} + \Gamma^{\beta}_{\mu\alpha}\Gamma^{\alpha}_{\nu\beta},$$

$$S_{\mu\nu} = \frac{\partial \Gamma^{\tau}_{\tau\mu}}{\partial x_\nu} - \Gamma^{\alpha}_{\mu\nu}\Gamma^{\tau}_{\tau\alpha}.$$

And the contraction components from the Riemann tensor,

$$\Gamma^{\tau}_{\tau\mu} = \frac{\partial \log\sqrt{-g}}{\partial x_\mu}, \Gamma^{\tau}_{\tau\alpha} = \frac{\partial \log\sqrt{-g}}{\partial x_\alpha}.$$

If one comes back to equation (18a) and chooses the coordinates so that $\sqrt{-g} = 1$, then $S_{\mu\nu}$ vanishes, and the Einstein tensor reduces to the Ricci tensor $R_{\mu\nu}$. Einstein noted, "I will therefore give below all relations in the simplified form, which this specialization of the choice of coordinates brings with it. It will then be easy to access to the *generally* covariant equations, if this seems desirable in a special case".[152]

Once postulating the $\sqrt{-g} = 1$, Einstein put aside his 1914 review article and he now based his theory on his November 1915 works, and he returned to his November 4, 1915 paper.

## 2.12 Theory Valid for $\sqrt{-g} = 1$

Einstein arrived at Part C, the theory of the gravitational field. In section §13 Einstein was occupied with the "Equations of motion of a material point in the gravitational field". In his November 4 1915 paper, Einstein arrived at the components of the gravitational field, equation (15a) of that paper: [153]

$$(45)\ \Gamma^{\tau}_{\mu\nu} = -\left\{ {\mu\nu \atop \tau} \right\}.$$

He then wrote in that paper the geodesic equations (15b), written here with Einstein's summation convention: [154]

$$(46)\ \frac{d^2x_\tau}{ds^2} = \Gamma^{\tau}_{\mu\nu}\frac{dx_\mu}{ds}\frac{dx_\nu}{ds}.$$

Consider a freely moving body not subject to external forces, and a local coordinate system $K_0$, from the point of view of the general theory of relativity, in a part of the four-dimensional space. In this local coordinate system the $g_{\mu\nu}$ have the special constant values given by equation (4). Thus with respect to $K_0$ the body moves in a straight line and in a uniform motion exactly as it would have moved with respect to an inertial system according to the special theory of relativity. [155]

Consider another (accelerated) coordinate system $K_1$. Recall the definition in section §2 of the equivalence principle. Therefore, the accelerated system $K_1$ is exchanged with an equivalent system in a gravitational field, which generates the acceleration. If we observe the motion of the body from this system $K_1$, then the body observed from $K_1$ moves in a gravitational field. [156]

With respect to $K_0$ the law of motion corresponds to a four-dimensional straight line, which is therefore a geodesic line. Since the geodesic line is defined independently of the reference system, its equations will also be the equation of motion of the material point with respect to $K_1$. With equations (45) the equations of the motion of the point with respect to $K_1$ become (46). If the geodesic line is an invariant then these equations define the motion of the point in a gravitational field in the case when there is no reference system $K_0$, with respect to which the special theory of relativity is valid. [157] If the components (45) vanish, then the point moves uniformly in a straight line. These quantities in (45) therefore condition the deviation of the motion from uniform motion, and Einstein called them, as he had done in November 1915, the components of the gravitational field.

In section §14 "The field equations of Gravitation in the Absence of Matter", Einstein merged among three papers: the November 11, November 18, and November 25 1915 papers. Einstein began the section with the distinction between "gravitational field" and "matter": we denote everything but the gravitational field as matter. Matter thus includes also the electromagnetic field.[158] Einstein started with the gravitational field and afterwards he was occupied with matter: the energy tensor, and Maxwell's equations in tensorial form.[159]

Einstein began with the field equations of gravitation *in the absence of matter*. He had already written the vacuum field equations in his November 18 paper. The gravitational field of the sun (in vacuum) satisfies the following field equations with respect to the special reference system $\sqrt{-g} = 1$ :[160]

$$(47)\quad \frac{\partial \Gamma^{\sigma}_{\mu\nu}}{\partial x_{\alpha}} + \Gamma^{\sigma}_{\mu\beta}\Gamma^{\beta}_{\nu\alpha} = 0,$$

$$\sqrt{-g} = 1.$$

In the November 25 paper Einstein wrote that the ten generally covariant equations of the gravitational field in the absence of matter are obtained by setting: [161] $G_{\mu\nu} = 0$. Subsequently Einstein wrote (47).

In 1916, Einstein first required the satisfaction of these equations in the special case of the special theory of relativity, in which the $g_{\mu\nu}$ have the certain constant values (4). Let this be the case in a certain finite area with respect to the coordinate system $K_0$. Then all the components of the Riemann tensor $B^{\rho}_{\mu\sigma\tau}$ [defined in (43)] vanish. For this finite area they also vanish for any other coordinate system. [162]

The equations of the matter free (vacuum) gravitational field must be satisfied anyway even if all the components of $B^{\rho}_{\mu\sigma\tau}$ vanish. However, this allows us to choose a coordinate system in which the gravitational field generated by a material point might be transformed away; i.e., it can be transformed to the case of *constant* $g_{\mu\nu}$.[163] Einstein explained that if we want to find the field equations of gravitation in the absence of matter, for matter-free *gravitational field*, so that the above case will not happen, we require that the symmetrical tensor $B_{\mu\nu}$, the Einstein tensor $G_{\mu\nu}$ (44), derived from the Riemann tensor $B^{\rho}_{\mu\sigma\tau}$ , should vanish. [164]

We obtain ten equations for the ten quantities $g_{\mu\nu}$, which are satisfied in the absence of matter, in the special case of the vanishing of all $G_{\mu\nu}$. With the choice of a coordinate system with respect to which $\sqrt{-g} = 1$ is valid, and taking into consideration (44), the equations for the matter free field are (47) .[165]

Einstein then wrote about equations (47),

"These equations proceed in a purely mathematical way from the requirement of the general theory of relativity, and give us in combination with the equations of motion (46), to a first approximation, Newton's law of attraction, and to a second approximation the explanation discovered by Leverior (as it remains after corrections after the perturbation have been made) for the perihelion motion of Mercury, which must, in my opinion, convince the physical correctness of the theory".[166]

In section §21 Einstein demonstrated that (47) gives us in combination with (46) Poisson's equation. In section §22 he showed that (47) gives us to second approximation the explanation for the precession of the perihelion of Mercury. Yet there was still no experimental proof verifying a prediction made by Einstein himself; namely, no experimental verification of a result derived from Einstein's new theory. The precession of the perihelion of Mercury was an unexplained phenomenon that bothered scientists even before Einstein had advanced his general theory of relativity. From the empirical point of view, mathematical conformity (as much as generally covariance was beautiful and promising) and an elegant theory were not yet enough to win over full success in 1916. Einstein was in the need of an empirical verification of a prediction of his own theory.

The next mission in the 1916 paper was achieved in section §15, "The Hamiltonian Function for the Gravitational Field, Law of Momentum and Energy".[167] In the manuscript of the 1916 paper Einstein wrote first "Impuls-Energie zu", he crossed out these words and then he wrote the above title.[168] Perhaps he thought to start with the conservation of energy momentum; and this signifies as shown in detail below that this section is intimately related to Einstein's calculations from his November 4, 1915 paper – calculations which are concerned with the momentum-energy conservation law.

After presenting new field equations in November 4 1915, Einstein demonstrated that these could be brought into Hamiltonian form. With this demonstration he tried to show that his field equations satisfy the conservation laws. However, the November 4 field equations were not yet casted in the final form; this was true especially for the right-hand term of the stress-energy tensor; and therefore the Hamiltonian form of the November 4 field equations led to the problematic equation (21a) of November 4.
In 1916 Einstein reproduced the procedure from his November 4 paper, but redefined the right hand source term in the November 4, 1915 equations. By doing so Einstein was led to the November 25, 1915 field equations valid in a coordinate system in which $\sqrt{-g} = 1$.[169]

In the 1916 paper Einstein corrected his November 4 1915 derivation and this led him straight to the November 25 field equations for a coordinate system in which $\sqrt{-g} = 1$. This is demonstrated now.

In section §15 Einstein showed that the field equations (47) correspond to the laws of momentum and energy.

Einstein first considered the November 4, 1915 Hamiltonian H, the second term of (47) multiplied with $g^{\mu\nu}$:[170]

$$H = g^{\mu\nu}\Gamma^{\alpha}_{\mu\beta}\Gamma^{\beta}_{\nu\alpha}.$$

Hence, obtaining the field equation from a variation principle, the action and Hamiltonian are,[171]

$$(47a)\ \delta\{\int H d\tau\} = 0, \quad H = g^{\mu\nu}\Gamma^{\alpha}_{\mu\beta}\Gamma^{\beta}_{\nu\alpha}, \ \sqrt{-g} = 1.$$

On the boundary of the finite four dimensional region of integration, the variation vanishes.

**First step: (47a) is equivalent to (47) – equations (48)**

Einstein's first step was to show that (47a) is equivalent to (47).[172]
For this purpose he regarded H as a function of $g^{\mu\nu}$ and defined:

$$g^{\mu\nu}_{\sigma}\left(= \frac{\partial g^{\mu\nu}}{\partial x_{\sigma}}\right).$$

Then the variation gives,

$$\delta H = \Gamma^{\alpha}_{\mu\beta}\Gamma^{\beta}_{\nu\alpha}\delta g^{\mu\nu} + 2g^{\mu\nu}\Gamma^{\alpha}_{\mu\beta}\delta\Gamma^{\beta}_{\nu\alpha} = -\Gamma^{\alpha}_{\mu\beta}\Gamma^{\beta}_{\nu\alpha}\delta g^{\mu\nu} + 2\Gamma^{\alpha}_{\mu\beta}\delta\left(g^{\mu\nu}\Gamma^{\beta}_{\nu\alpha}\right)$$

But the last term on the right-hand side in the round brackets is equal to,

$$\delta\left(g^{\mu\nu}\Gamma^{\beta}_{\nu\alpha}\right) = -\tfrac{1}{2}\delta\left[g^{\mu\nu}g^{\beta\lambda}\left(\frac{\partial g_{\nu\lambda}}{\partial x_\alpha} + \frac{\partial g_{\alpha\lambda}}{\partial x_\nu} - \frac{\partial g_{\alpha\nu}}{\partial x_\lambda}\right)\right].$$

Because of (45),

$$\Gamma^{\tau}_{\mu\nu} = -\frac{1}{2}g^{\tau\alpha}\left(\frac{\partial g_{\mu\alpha}}{\partial x_\nu} + \frac{\partial g_{\nu\alpha}}{\partial x_\mu} - \frac{\partial g_{\mu\nu}}{\partial x_\alpha}\right).$$

We first write the components of the gravitational field in the equation for δH using the above equations. Then the last two terms in the round brackets of $\delta\left(g^{\mu\nu}\Gamma^{\beta}_{\nu\alpha}\right)$ cancel each other in the terms $\Gamma^{\alpha}_{\mu\beta}$ and $\Gamma^{\beta}_{\nu\alpha}$ of δH. After cancelation of these terms we are left only with the first term in the expression of $\delta\left(g^{\mu\nu}\Gamma^{\beta}_{\nu\alpha}\right)$. Taking (31) into account, then this term can be rewritten in the following form,

$$\delta g^{\mu\nu}g^{\beta\lambda}\frac{\partial g_{\nu\lambda}}{\partial x_\alpha} = -\delta\frac{\partial g_{\mu\beta}}{\partial x_\alpha} = -\delta g^{\mu\beta}_{\alpha}.$$

So finally we are left only with the following terms in the equation for δH,

$$\delta H = -\Gamma^{\alpha}_{\mu\beta}\Gamma^{\beta}_{\nu\alpha}\delta g^{\mu\nu} - \Gamma^{\alpha}_{\mu\beta}\delta g^{\mu\beta}_{\alpha}.$$

This is the variation. From this Einstein obtained, [173]

$$\textbf{(48)}\ \frac{\partial H}{\partial g^{\mu\nu}} = -\Gamma^{\alpha}_{\mu\beta}\Gamma^{\beta}_{\nu\alpha}, \qquad \frac{\partial H}{\partial g^{\mu\nu}_{\sigma}} = \Gamma^{\sigma}_{\mu\nu}.$$

Carrying the variation in (47a) [$g_{\mu\nu}$ in (47a): $g^{\mu\nu}_{\alpha} = \frac{\partial g^{\mu\nu}}{\partial x_\alpha}$],

$$\textbf{(47b)}\ \frac{\partial}{\partial x_\alpha}\left(\frac{\partial H}{\partial g^{\mu\beta}_{\alpha}}\right) - \frac{\partial H}{\partial g^{\mu\nu}} = 0.$$

Inserting (48) into (47b) gives the field equations for the matter-free gravitational field (47).

The left hand side of equation (47b) is equation (18) from the November 4, 1915 paper.[174] The 1915 equation (18) contained the sources on the right-hand side:[175]

$$\frac{\partial}{\partial x_\alpha}\left(\frac{\partial \mathrm{H}}{\partial \mathrm{g}_\alpha^{\mu\beta}}\right)-\frac{\partial \mathrm{H}}{\partial \mathrm{g}^{\mu\nu}}=-\kappa \mathrm{T}_{\mu\nu}.$$

In the November 4, 1915 paper Einstein also wrote equations (48) as equations (19) and (19a), and inserted them into the above equation in order to obtain his November 4, 1915 field equations,

$$\frac{\partial \Gamma_{\mu\nu}^{\sigma}}{\partial x_\alpha}+\Gamma_{\mu\beta}^{\sigma}\Gamma_{\nu\alpha}^{\beta}=-\kappa T_{\mu\nu}$$

in non-Hamiltonian formalism (written here with Einstein's summation convention).[176]

**Second step: rewriting (47b) in the form of (51)**

Einstein could thus use (47b) and (48) to show that the vacuum field equations correspond to the laws of momentum and energy. The first step was writing (47b) in a new form. Einstein obtained this result by exactly following his procedure from the November 4, 1915 paper. There Einstein started by multiplying (47b) [with the sources term then known to Einstein] by $\mathrm{g}_\sigma^{\mu\nu}$ with summation over the indices μ and ν:[177]

$$\mathrm{g}_\sigma^{\mu\nu}\frac{\partial}{\partial x_\alpha}\left(\frac{\partial \mathrm{H}}{\partial \mathrm{g}_\alpha^{\mu\nu}}\right)=\frac{\partial}{\partial x_\alpha}\left(\mathrm{g}_\sigma^{\mu\nu}\frac{\partial \mathrm{H}}{\partial \mathrm{g}_\alpha^{\mu\nu}}\right)-\frac{\partial \mathrm{H}}{\partial \mathrm{g}_\alpha^{\mu\nu}}\frac{\partial \mathrm{g}_\alpha^{\mu\nu}}{\partial x_\sigma}$$

Einstein used here the definition, $\frac{\partial \mathrm{g}_\sigma^{\mu\nu}}{\partial x_\alpha}=\frac{\partial \mathrm{g}_\alpha^{\mu\nu}}{\partial x_\sigma}$

He obtained in the November 4, 1915 paper the following end result: [178]

$$\text{(November 4)}\quad \frac{\partial}{\partial x_\alpha}\left(\mathrm{g}_\sigma^{\mu\nu}\frac{\partial \mathrm{H}}{\partial \mathrm{g}_\alpha^{\mu\nu}}\right)-\frac{\partial \mathrm{H}}{\partial x_\sigma}=-\kappa \mathrm{T}_{\mu\nu}\mathrm{g}_\sigma^{\mu\nu},$$

And the 1916 equation,

$$\frac{\partial}{\partial x_\alpha}\left(g_\sigma^{\mu\nu}\frac{\partial H}{\partial g_\alpha^{\mu\nu}}\right)-\frac{\partial H}{\partial x_\sigma}=0.$$

In the November 4 paper the sources entered on the right hand side; all the above equations are written here with Einstein's 1916 summation convention, as it appears in the 1916 paper. However, now Einstein added the sources but corrected them.

In the manuscript of the 1916 paper, Einstein wrote next to the above highlighted equation, (49) and crossed it out. Therefore, this equation was intended to be (49) and Einstein regretted in the last minute.[179] He decided to number the equation below (49) very likely because of the factor $-2\kappa t_\sigma^\alpha$ that entered into this equation.

In November 4 Einstein defined the energy tensor of the gravitational field in equation (20a) of the November 4, 1915 paper (omitting the summation from the first term on the right-hand side of the November 4 equation) – in 1915 and in 1916 Einstein wrote $\chi$ instead of $\kappa$, but in this chapter we shall write $\kappa$: [180]

$$(49)\quad -2\kappa t_\sigma^\alpha = g_\sigma^{\mu\nu}\frac{\partial H}{\partial g_\alpha^{\mu\nu}} - \delta_\sigma^\alpha H.$$

To this November 4 equation Einstein added in 1916 another equation:

$$\frac{\partial t_\sigma^\alpha}{\partial x_\alpha} = 0.$$

Both these equations consisted equations (49).

In the November 4, 1915 paper Einstein said that using the second of equations (48), the components of the gravitational field, then the first equation of (49) could also be written in the following form (20b),[181]

$$(50)\quad \kappa t_\sigma^\alpha = \frac{1}{2}\delta_\sigma^\alpha g^{\mu\nu}\Gamma_{\mu\beta}^\alpha\Gamma_{\nu\alpha}^\beta - g^{\mu\nu}\Gamma_{\mu\beta}^\alpha\Gamma_{\nu\sigma}^\beta.$$

In the manuscript of the 1916 paper Einstein wrote equation (50) twice. Above this equation Einstein wrote exactly the same equation, but he thought it was incomprehensible, because of his hesitations over the indices. Hence finally he crossed out the additional equation and left equation (50).[182]

Einstein used equation (34) $\frac{\partial g_{\mu\nu}}{\partial x_\sigma} = -g^{\mu\tau}\begin{Bmatrix}\tau\sigma \\ \nu\end{Bmatrix} - g^{\nu\tau}\begin{Bmatrix}\tau\sigma \\ \mu\end{Bmatrix}$ in equation (50). [183]

In 1916 Einstein said that (49) applies to all systems of coordinates for which $\sqrt{-g} = 1$. Indeed this is true for Einstein's procedure from November 4, 1915. Einstein summarized regarding (50), "This equation expresses the law of conservation of momentum and energy for the gravitational field".[184] The quantities $t_\sigma^\alpha$ are the "energy components" of the gravitational field.

In the November 4, 1915 paper Einstein multiplied (47) with the sources by $g^{\nu\sigma}$ (then known to Einstein on November 4):[185]

$$(I) \quad g^{\nu\sigma}\frac{\partial \Gamma_{\mu\nu}^{\alpha}}{\partial x_\alpha} - g^{\nu\sigma}\Gamma_{\nu\beta}^{\alpha}\Gamma_{\nu\alpha}^{\beta\lambda} = -\kappa g^{\nu\sigma}T_{\mu\nu} = -\kappa T_\mu^\tau.$$

The second term on the left-hand side above is the second term on the right-hand side of (50), and so Einstein combined between the two and obtained the following equation, [186]

$$(II) \quad \frac{\partial}{\partial x_\alpha}\left(g^{\sigma\beta}\Gamma_{\mu\beta}^{\alpha}\right) - \frac{1}{2}\delta_\sigma^\alpha g^{\mu\nu}\Gamma_{\mu\beta}^{\lambda}\Gamma_{\nu\lambda}^{\beta} = -\kappa\left(T_\mu^\tau + t_\mu^\lambda\right).$$

After this equation Einstein was led to the problematic equation (21a) from his November 4, 1915 paper. [187]

In 1916 Einstein corrected the above equation in the following way (and by thus prevented the above problem). He followed exactly the above November 4 procedure and multiplied (47) by $g^{\nu\sigma}$. Einstein wrote: [188]

$$g^{\nu\sigma}\frac{\partial \Gamma_{\mu\nu}^{\alpha}}{\partial x_\alpha} = \frac{\partial}{\partial x_\alpha}\left(g^{\sigma\beta}\Gamma_{\mu\beta}^{\alpha}\right) - \frac{\partial g^{\nu\sigma}}{\partial x_\alpha}\Gamma_{\mu\nu}^{\alpha},$$

And because of relation (34), which also led to (50),

$$\frac{\partial g^{\nu\sigma}}{\partial x_\alpha} = -g^{\nu\beta}\Gamma_{\alpha\beta}^{\sigma}\Gamma_{\mu\nu}^{\alpha} - g^{\sigma\beta}\Gamma_{\beta\alpha}^{\nu}\Gamma_{\mu\nu}^{\alpha},$$

The above equation is written in the following form,[189]

$$\frac{\partial}{\partial x_\alpha}\left(g^{\sigma\beta}\Gamma^{\alpha}_{\mu\beta}\right) - g^{\nu\beta}\Gamma^{\sigma}_{\alpha\beta}\Gamma^{\alpha}_{\mu\nu} - g^{\sigma\beta}\Gamma^{\nu}_{\beta\alpha}\Gamma^{\alpha}_{\mu\nu}.$$

Einstein changed the symbols of the summation indices,[190]

$$\frac{\partial}{\partial x_\alpha}\left(g^{\sigma\beta}\Gamma^{\alpha}_{\mu\beta}\right) - g^{\gamma\delta}\Gamma^{\sigma}_{\gamma\beta}\Gamma^{\beta}_{\delta\mu} - g^{\nu\sigma}\Gamma^{\alpha}_{\mu\beta}\Gamma^{\beta}_{\nu\alpha}.$$

However, at this stage Einstein got mixed up with the indices…. and he switched between the second term and the third term in this equation. This is evident from reading the manuscript of the 1916 paper and looking at the following equation in this manuscript,

$$\frac{\partial}{\partial x_\alpha}\left(g^{\sigma\beta}\Gamma^{\alpha}_{\mu\beta}\right) - g^{\nu\sigma}\Gamma^{\alpha}_{\mu\beta}\Gamma^{\beta}_{\nu\alpha} - g^{\gamma\delta}\Gamma^{\sigma}_{\gamma\beta}\Gamma^{\beta}_{\delta\mu}.$$

Einstein soon realized that he made a mistake, and he drew a round arc above the second term indicating to the editor that one should switch between the two.[191]

The third term in the printed version of the above equation in the new indices then cancels itself with the second term of the field equations (47). We are thus left with the first and the second term,

$$\frac{\partial}{\partial \mathbf{x}_\alpha}\left(\boldsymbol{g}^{\sigma\beta}\boldsymbol{\Gamma}^{\alpha}_{\mu\beta}\right) - \boldsymbol{g}^{\gamma\delta}\boldsymbol{\Gamma}^{\sigma}_{\gamma\beta}\boldsymbol{\Gamma}^{\beta}_{\delta\mu} = \mathbf{0}.$$

This is the equivalent equation to the November 4, 1915 highlighted equation written above (I), but for matter-free gravitational field.

Using equation (50) we arrive at the equivalent to the second above highlighted equation of the November 4, 1915 paper, equation (II), but valid for matter-free gravitational fields,

$$\frac{\partial}{\partial \mathbf{x}_\alpha}\left(\boldsymbol{g}^{\sigma\beta}\boldsymbol{\Gamma}^{\alpha}_{\mu\beta}\right) - \frac{1}{2}\boldsymbol{\delta}^{\alpha}_{\sigma}\boldsymbol{g}^{\mu\nu}\boldsymbol{\Gamma}^{\lambda}_{\mu\beta}\boldsymbol{\Gamma}^{\beta}_{\nu\lambda} = -\boldsymbol{\kappa}\mathbf{t}^{\alpha}_{\sigma}.$$

If the second term on the right hand side, $g^{\mu\nu}\Gamma^{\lambda}_{\mu\beta}\Gamma^{\beta}_{\nu\lambda} = \kappa t$, then the above equation is rewritten in the following form, [192]

$$(51)\ \frac{\partial}{\partial x_{\alpha}}\left(g^{\sigma\beta}\Gamma^{\alpha}_{\mu\beta}\right) = -\kappa\left(t^{\sigma}_{\mu} - \frac{1}{2}\delta^{\sigma}_{\mu}t\right), \qquad \sqrt{-g} = 1.$$

Einstein then moved on to the next section, section §16 "The General Form of the Field Equations of Gravitation".[193]

How could equations (51) solve the problem of the November 4 paper? Let us come back for a moment to the above field equations (II) from November 4, written here with the summation convention,[194]

$$(\mathrm{II})\ \frac{\partial}{\partial x_{\alpha}}\left(g^{\sigma\beta}\Gamma^{\alpha}_{\mu\beta}\right) - \frac{1}{2}\delta^{\alpha}_{\mu}\kappa t = -\kappa\left(T^{\sigma}_{\mu} + t^{\sigma}_{\mu}\right).$$

Instead of this equation Einstein wrote the following, [195]

$$(52)\ \frac{\partial}{\partial x_{\alpha}}\left(g^{\sigma\beta}\Gamma^{\alpha}_{\mu\beta}\right) = -\kappa\left[\left(t^{\sigma}_{\mu} + T^{\sigma}_{\mu}\right) - \frac{1}{2}\delta^{\alpha}_{\mu}(t + \mathbf{T})\right], \qquad \sqrt{-g} = 1.$$

Einstein added to the November 4, 1915 equation (II) one single term: $-\frac{1}{2}\delta^{\alpha}_{\mu}T$.

This additional term leads to the famous field equations of November 25 valid for coordinate system $\sqrt{-g} = 1$. It is perfectly reasonable to add this term for reasons of symmetry with relation to the second term on the right hand of equation (52). This term is $\mathrm{T} = T^{\mu}_{\mu}$ and it is Laue's scalar.

A glance at Einstein's correspondence with Paul Ehrenfest during January 1916 supplies an answer to the second question: Einstein laid down the above derivation to Ehrenfest in the letters he had sent him during January 1916 as a reply to the latter's quarries: equations (47) and (53) below,[196] equations (50) and the equations (derivation) leading to equation (51) and then Einstein wrote to Ehrenfest equation (52); designated in the letter to Ehrenfest as equation (8). [197]

Einstein wrote Ehrenfest,

"This equation is interesting, because it shows that the source of the gravitation lines is determined solely by the sum $T_\sigma^\nu + t_\sigma^\nu$, as indeed it must be expected".[198]

Einstein also wrote Ehrenfest a few days earlier about the "'inevitability' for the additional term – 1/2 $g_{im}T$"[199].

At the end of the letter Einstein asked Ehrenfest to send him back the pages of these formulas, because nowhere else did he have these equations so nicely written in one place. And these pages formed the basis for his derivation, including equation (52) [(8)].

Equations (52) represent the required general field equations of gravitation ($\sqrt{-g} = 1$) in mixed form. From the above field equations Einstein arrived at his November 25 field equations normalized for $\sqrt{-g} = 1$. It is very important to remember that (52) hold for a coordinate system in which $\sqrt{-g} = 1$, and therefore it is an extension of the November 4 equivalent equation.

From (52),

$$\frac{\partial}{\partial x_\alpha}\left(g^{\sigma\beta}\Gamma_{\mu\beta}^{\alpha}\right) = -\boldsymbol{\kappa t_\mu^\sigma} + \frac{\mathbf{1}}{\mathbf{2}}\boldsymbol{\kappa\delta_\mu^\alpha t} - \kappa T_\mu^\sigma - \frac{1}{2}\kappa\delta_\mu^\alpha T$$

And according to (51) and (47), as Einstein showed to Ehrenfest on January 24,[200]

$$g^{\nu\sigma}\left(\frac{\partial\Gamma_{\mu\nu}^{\alpha}}{\partial x_\alpha} + \Gamma_{\mu\beta}^{\alpha}\Gamma_{\nu\alpha}^{\beta}\right) - \boldsymbol{\kappa t_\mu^\sigma} + \frac{\mathbf{1}}{\mathbf{2}}\boldsymbol{\kappa\delta_\mu^\alpha t} - \kappa\left(T_\mu^\sigma - \frac{1}{2}\delta_\mu^\alpha T\right).$$

Multiplying by $g_{\mu\nu}$ we thus get,[201]

$$\textbf{(53)}\ \boldsymbol{\frac{\partial\Gamma_{\mu\nu}^{\alpha}}{\partial x_\alpha} + \Gamma_{\mu\beta}^{\alpha}\Gamma_{\nu\alpha}^{\beta} - \kappa\left(T_{\mu\nu} - \frac{1}{2}g_{\mu\nu}T\right)}.$$

$$\boldsymbol{\sqrt{-g} = 1}.$$

Einstein later obtained this equation by correcting the November 4, 1915 field equations in the following way. The November 4 field equations are $R_{\mu\nu} = -\kappa T_{\mu\nu}$.

Einstein wrote instead, $R_{\mu\nu} - 1/2g_{\mu\nu}R = -\kappa T_{\mu\nu}$. Multiplying this equation by $g^{\mu\nu}$, and summing over μ and ν, gives, $R = \kappa g^{\mu\nu}T_{\mu\nu} = \kappa T$ (where, $g^{\mu\nu}g_{\mu\nu} = 4$). Putting this value in the above equation gives (53). [202]

Einstein then said, [203]

"It must be admitted that this introduction of the energy tensor of matter is not justified by the relativity postulate alone; therefore we have deduced it from the requirement that the energy of the gravitational field shall act gravitationally in the same way as any other kind of energy. The strongest reason for the choice of these equations, however, lies in their consequences, that equations (49) […] are valid for the components of total-energy conservation laws (of momentum and energy)".

Einstein was going to show this in section §17. He contracted equation (52), rearranged terms using (29) and (31), and obtained the desired result, [204]

$$(56)\ \frac{\partial\left(t_{\mu}^{\sigma} + T_{\mu}^{\sigma}\right)}{\partial x_{\alpha}} = 0.$$

Einstein summarized, "Thus it is therefore apparent from our field equations of gravitation that the laws of conservation of momentum and energy are satisfied".[205]

In section §18 Einstein wrote the laws of momentum and energy for matter, as a consequence of the field equations. Einstein obtained from (53) and (56), [206]

$$(57)\ \frac{\partial T_{\sigma}^{\alpha}}{\partial x_{\alpha}} + \frac{1}{2}\frac{\partial g^{\mu\nu}}{\partial x_{\sigma}} T_{\mu\nu} = 0.$$

This equation represents the vanishing of divergence of the material energy-tensor.

The second term on the left hand side shows that the laws of conservation of momentum and energy do not apply for matter alone, but that momentum and energy are transferred from the gravitational field to matter. Einstein demonstrated this by rewriting equation (57) in the following form, [207]

$$(57a)\ \frac{\partial T_{\sigma}^{\alpha}}{\partial x_{\alpha}} = -\Gamma_{\sigma\beta}^{\alpha}\Gamma_{\alpha}^{\beta}.$$

Einstein used equation (41) for the divergence of a mixed tensor, [208]

$$\sqrt{-g}\mathrm{A}_\mu = \frac{\partial\left(\sqrt{-g}A^\sigma_\mu\right)}{\partial \mathrm{x}_\sigma} - \left\{ {\sigma\mu \atop \tau} \right\} \sqrt{-g}A^\sigma_\tau .$$

Since $\sqrt{-g} = 1$,

$$\mathrm{A}_\mu = \frac{\partial A^\sigma_\mu}{\partial \mathrm{x}_\sigma} - \Gamma^\alpha_{\sigma\beta} A^\sigma_\tau .$$

(57a) is thus dependent on the choice $\sqrt{-g} = 1$.

The term on the right hand side of equation (57a) is the second term in equation (47), the field equations for matter-free gravitational field. Thus combining between the two we obtain,

$$\frac{\boldsymbol{\partial T^\alpha_\sigma}}{\boldsymbol{\partial \mathrm{x}_\alpha}} = \frac{\boldsymbol{\partial \Gamma^\alpha_{\mu\nu}}}{\boldsymbol{\partial \mathrm{x}_\alpha}}.$$

The right-hand side of this equation expresses the energetic action of the gravitational field on matter. Therefore the field equations of gravitation, said Einstein, give the equations of material processes completely, if these later are characterized by four differential equations independent of one another. For this Einstein referred to Hilbert's paper from 1915. [209]

**2.13 Matter: Energy-Momentum Tensor**

Einstein arrived at section D, "Material Processes". The Einstein-Ricci tensor $B_{\mu\nu}$, equation (44) included the definition of the gravitational field. Recall that at the beginning of section §14 Einstein demarcated "gravitation" from "matter".[210] Matter was represented by the energy tensor. Einstein also included in "matter" the electromagnetic field. Einstein started with the former tensor, the energy tensor. Subsequently he derived the electromagnetic equations in tensorial form.

In section §19 Einstein defined the contravariant stress-energy tensor in terms of the pressure and the density of a fluid (a model for a flow of matter, dust),[211]

$$(58)\ T^{\alpha\beta} = -g^{\alpha\beta}p + \rho \frac{dx_\alpha}{ds}\frac{dx_\beta}{ds},$$

And the covariant energy-tensor of the fluid is,

$$(58a)\ T_{\mu\nu} = -g_{\mu\nu}p + g_{\mu\sigma}\frac{dx_\alpha}{ds}g_{\mu\beta}\frac{dx_\beta}{ds}\rho.$$

Einstein had already written these equations in the 1914 review article using tensor density – V-tensors. With Einstein's summation convention, his 1914 equation for the mixed energy-tensor of matter is given by,[212]

$$\mathfrak{T}_\sigma^\nu = \rho\sqrt{-g}\frac{dx_\nu}{ds}g_{\sigma\mu}\frac{dx_\mu}{ds}.$$

Thus according to Einstein's field equations (53), the energy tensor was calculated from the $g_{\mu\nu}$, and this energy tensor was given by,[213]

$$(58)\ \boldsymbol{T^{\alpha\beta} = \rho\frac{dx_\alpha}{ds}\frac{dx_\beta}{ds}},$$

for a flow of pressureless matter (dust).

**2.14 Matter: Electromagnetic Equations**

In section §20 Einstein arrived at Maxwell's electromagnetic equations. Einstein corrected section §11 of his 1914 paper "The Electromagnetic Equations".[214]

Einstein started from the components of the electromagnetic field, $F_{\rho\sigma}$, which are related with the electromagnetic potential φ(φ, A), and wrote them according to the formula for the curl of the covariant vector:[215]

$$(59)\ F_{\rho\sigma} = \frac{\partial\varphi_\rho}{\partial x_\sigma} - \frac{\partial\varphi_\sigma}{\partial x_\rho}.$$

From this, the following system of equations follows,[216]

$$(60)\ \boldsymbol{\frac{\partial F_{\rho\sigma}}{\partial x_\tau} + \frac{\partial F_{\sigma\tau}}{\partial x_\rho} + \frac{\partial F_{\tau\rho}}{\partial x_\sigma} = 0}.$$

This is writing Maxwell's equations in a tensorial form,

$$div\boldsymbol{H} = 0, \text{or}\ \frac{\partial H_x}{\partial x} + \frac{\partial H_y}{\partial y} + \frac{\partial H_z}{\partial z} = 0.$$

$$curl\boldsymbol{E} = -\frac{\partial \boldsymbol{H}}{\partial t}, \text{or } \frac{\partial H_x}{\partial x_4} = \frac{\partial F_{\mu\nu}}{\partial y} - \frac{\partial F_{\nu\tau}}{\partial z} \ldots$$

Both these equations consisted equation (60a) and equations (60b).[217]

Einstein introduced the contravariant vector $J^{\mu}$ of the electric current density in vacuum. He also wrote the components of the electromagnetic field in a contravariant form,[218]

$$(62)\ \boldsymbol{F^{\mu\nu} = g^{\mu\alpha} g^{\nu\beta} F_{\alpha\beta}}.$$

He used the expression of a divergence of a contravariant six vector (40),[219]

$A^{\alpha} = \frac{1}{\sqrt{-g}} \frac{\partial\left(\sqrt{-g}A^{\alpha\beta}\right)}{\partial x_{\beta}}$ Since $\sqrt{-g} = 1$ in this expression, then $A^{\alpha} = \frac{\partial A^{\alpha\beta}}{\partial x_{\beta}}$.

Thus accordingly,[220]

$$(63)\ \frac{\partial \boldsymbol{F^{\mu\nu}}}{\partial \boldsymbol{x_{\nu}}} = \mathbf{J}^{\mu}.$$

This is $\frac{\partial E}{\partial t} = -J$.

In the case of the special theory of relativity $J^4 = \rho$.

Then,

$$div\frac{\partial \boldsymbol{E}}{\partial t} = \frac{\partial \rho}{\partial t}.$$

This leads to,[221]

$$(63a)\ curl\boldsymbol{H} - \frac{\partial \boldsymbol{E}}{\partial t} = J, div\boldsymbol{E} = \rho.$$

The highlighted equations are the generalization of Maxwell's equations for free space, with respect to the choice of coordinates in which $\sqrt{-g} = 1$.

Einstein next formed the inner product,[222]

(65) $\kappa_\sigma = F_{\sigma\mu}J^\mu$,

and defined $\kappa_\sigma$ as a covariant four vector, the components of which are the energy transferred from the electric masses to the electromagnetic field per unit of time per unit volume. If the electric masses are free, i.e., are only under the influence of the electromagnetic field, then $\kappa_\sigma$ will vanish.

Einstein was about to obtain the energy components of the electromagnetic field, $T_\sigma^\nu$. He combined between equations (63) and (65) and obtained two equations similar to (57),[223]

$$(66)\ \kappa_\sigma = \frac{\partial T_\sigma^\nu}{\partial x_\nu} - \frac{1}{2} g^{\tau\mu} \frac{\partial g_{\mu\nu}}{\partial x_\sigma} T_\tau^\nu,$$

(these equations are exactly similar to equations (57) if $\kappa_\sigma$ vanishes), where,

$$(66a)\ T_\sigma^\nu = -F_{\sigma\alpha}F^{\nu\alpha} + \frac{1}{4}\delta_\sigma^\nu F_{\alpha\beta}F^{\alpha\beta}.$$

Einstein concluded that the above $T_\sigma^\nu$ are the energy components of the electromagnetic field.

Einstein wrote in his November 11 paper, "There are even quite a few, who hope to reduce matter to purely electromagnetic processes, which, however, would have to be done in a theory more completed than Maxwell's electrodynamics". The energy tensor of "matter" $T_\mu^\lambda$ has a scalar (trace) $\sum_\mu T_\mu^\mu$. It vanishes in an electromagnetic field. But it differs from zero for matter proper.[224] The above tensor $T_\sigma^\nu$ then has a "scalar" which is equal to zero.

It is almost as if Einstein went back to his November 11 paper. After equation (66a) Einstein even apologized for not developing in his paper generally covariant field equations, but only field equations covariant with respect to a coordinate system in which $\sqrt{-g} = 1$.

However, in the manuscript of the 1916 paper this apology appeared in a note which consisted page 40a (a sheet which was half written and the bottom half was empty). This sheet appeared in the manuscript *after* sheet (page) number 40 which consisted *the beginning of part E and section §21 dealing with Newton's theory*. In the manuscript Einstein mentioned that the note is an addition to part D and he signed both pages 40 and 40a on the top with a red *.[225]

The apology was the following: Einstein wrote that we have now derived the most general laws for the gravitational field and matter for a coordinate system, for which $\sqrt{-g} = 1$. By this we have achieved a considerable simplification of the formulas and calculations, without having to omit the requirement of general covariance: because our equations were found, through specialization of a coordinate system, from generally covariant equations. [226] Einstein had already said this before in the paper,[227] and now he simply apologized again! He still wondered whether (56) and the field equations (52) could be formulated – without assuming $\sqrt{-g} = 1$ – so that we would arrive at conservation of energy and momentum. Einstein then noted, "I have found that both are in fact the case". But he decided not to communicate these "comprehensive considerations", because they do not give anything objectively new.[228] This was the note added after equation (66a).

**2.15 Newtonian Limit**

Einstein now came to the final part of his paper, part E. This part was unnamed. It no more included material pertaining to the general theory of relativity. It was dedicated to applications and predictions of the theory that was presented in the previous parts. Next to the letter "E" Einstein started with section §21, "Newtonian Theory as a First Approximation".[229] This section was essential to the predictions that Einstein made afterwards.

In the manuscript of the 1916 paper, Einstein wrote a different title next to §21 and crossed it out: "Gesichtspunkte für die Aufstellung von näherungsweise gültigen Gesetzen". ("Considerations for the Establishment of Laws Applicable Approximately").[230] This title is interesting because it reflects a hesitation: Perhaps even before submitting the paper Einstein was not sure that he would be able to extract the Newtonian limit. This scenario is supported by the following finding. As seen in the manuscript of the 1916 paper, this section contains quite a few deletions of equations, and this seems to reflect Einstein's hesitations.

In the manuscript of the 1916 paper Einstein considered the motion of a material point according to equations (46). He then considered the case of the special theory of relativity. And then it appears that he hesitated, and added later (after writing the text) for clarity the following equations, $\frac{dx_1}{ds}, \frac{dx_2}{ds}, \frac{dx_3}{ds}$, which also appear in the printed version. Einstein concluded that any velocity can occur which *is less than the velocity of light*, and he wrote the following equation and it was crossed out,[231]

$$\left(\frac{dx_1}{ds}\right)^2 + \left(\frac{dx_2}{ds}\right)^2 + \left(\frac{dx_3}{ds}\right)^2 < 1.$$

*Einstein followed here special relativity*, but he deleted this equation.

Instead of this equation Einstein tried a new strategy, a velocity defined *in the sense of Euclidean geometry*, and he wrote the following equation,

$$[(I)] \qquad v = \sqrt{\frac{dx_1{}^2}{dx_4} + \frac{dx_2{}^2}{dx_4} + \frac{dx_3{}^2}{dx_4}},$$

and then the requirement that v < 1.

(Einstein would write this equation again towards the end of his paper when deriving the bending of light rays in a gravitational field).

Einstein again wrote the components,

$$[(II)] \qquad \frac{dx_1}{ds}, \frac{dx_2}{ds}, \frac{dx_3}{ds}$$

(Einstein would also write this equation when deriving the bending of light rays in a gravitational field).

After the above components Einstein tried once more *the old strategy of special relativity*, and wrote the following equation, but it was immediately crossed out, [232]

$$\frac{dx_1}{ds} \ll 1.$$

He said that if v is small as compared to the velocity of light, then the components above are treated as small quantities, while up to the second order quantities, $dx_4/ds$ is 1.[233] *Einstein also crossed out in the manuscript the term $dx_4/ds$. However, it appeared in print.*[234]

In the perihelion of Mercury November 18, 1915 paper, Einstein obtained the Newtonian equations of motion from the geodesic equation. Einstein wrote equations of motion for a point mass moving in the gravitational field of the sun to second order. He implemented the geodesic equation. Einstein said, from this equation we conclude that the Newtonian equation of motion is contained in it as a first approximation. He explained this further: when the speed of the planet is small with respect to the speed of light, then $dx_1$, $dx_2$, $dx_3$ are small as compared to $dx_4$. It follows from this that we get back the solutions, $g_{\rho\sigma}$, in the first order approximation, in which we always take on the right-hand side the condition $\sigma = \tau = 4$.[235] Einstein first calculated the components $\Gamma^{\tau}_{\rho\sigma}$ of the gravitational field of the sun to the first order approximation.

The first order solutions are substituted in the equations (45) of the components of the gravitational field of the sun. For the 44 component, Einstein obtained: [236]

$$\Gamma^{\sigma}_{44} = \Gamma^{4}_{4\sigma} = -\frac{\alpha}{2}\frac{x_\sigma}{r^3}.$$

Using the above equation Einstein then got: [237]

$$\frac{d^2 x_\nu}{ds^2} = \Gamma^{\nu}_{44} = -\frac{\alpha}{2}\frac{x_\nu}{r^3} \ (\nu = 1,2,3),\ \frac{d^2 x_4}{ds^2} = 0.$$

Einstein wrote that for the first approximation one can set $s = x_4$, that is, **s = t**; then the first three equations are exactly the Newtonian equations. [238]

In section §21 of the 1916 review article Einstein generalized this procedure. He wrote the above equation (with s = t), [239]

$$\frac{d^2 x_\tau}{dt^2} = \Gamma^{\tau}_{44}.$$

In the manuscript of the 1916 paper Einstein wrote above this equation the following equation and it was then crossed out,[240]

$$\frac{d^2x_\tau}{dt^2} = \begin{bmatrix}44\\1\end{bmatrix} + \begin{bmatrix}44\\2\end{bmatrix} + \begin{bmatrix}44\\3\end{bmatrix} - \begin{bmatrix}44\\4\end{bmatrix}.$$

After deleting this equation he settled on the equation below, which also appeared in the printed version,

$$\frac{d^2x_\tau}{dt^2} = -\begin{bmatrix}44\\4\end{bmatrix}.$$

The deleted equation above is interesting, because Einstein was already thinking about the field equations (53) and the approximation that would yield Poisson's equation, and he simply mixed between (46) and (53).

In the November 18, 1915 paper the gravitational field of the sun is static and spherically symmetric. In section §21 of the 1916 paper Einstein made the same assumption. The matter generating this field is slow compared to the velocity of light. And Einsteinw wrote,

$$\Gamma_{44}^{\tau} = \begin{bmatrix}44\\\tau\end{bmatrix} = \frac{1}{2}g_{44}\left(\frac{\partial g_{4\tau}}{\partial x_4} + \frac{\partial g_{4\tau}}{\partial x_4} - \frac{\partial g_{44}}{\partial x_\tau}\right) = -\frac{1}{2}\frac{\partial g_{44}}{\partial x_\tau}.$$

Therefore the above geodesic equation can be rewritten in the following form,[241]

$$(67)\ \frac{d^2x_\tau}{dt^2} = -\frac{1}{2}\frac{\partial g_{44}}{\partial x_\tau} \quad (\tau = 1,2,3).$$

This is the equation of motion of the material point according to Newton's theory, in which $g_{44}/2$ plays the part of the gravitational potential. Einstein had already got equation (67) in the November 18, 1915 paper. Einstein said that the remarkable thing with this result is that the fundamental tensor alone defines, to a first approximation, the motion of the material point.

### 2.16 Poisson's Equation

Einstein now turned to the field equations (53). In the approximation of the weak fields, the energy tensor of "matter" is almost exclusively defined by the density of matter, according to equation (58a).[242]

$$T_{44} = \frac{dx_4}{dt}\frac{dx_4}{dt}\rho = \rho = T.$$

**Thus the right hand side of (53) is:**

$$-\kappa\rho + \frac{1}{2}\kappa\rho = -\frac{1}{2}\boldsymbol{\kappa\rho}$$

As to the left-hand side of equation (53), the second term is a small quantity of second order. The first term yields,[243]

$$\frac{\partial}{\partial x_\tau}\Gamma^\tau_{\mu\nu} = \frac{\partial}{\partial x_1}\begin{bmatrix}\mu\nu\\1\end{bmatrix} + \frac{\partial}{\partial x_2}\begin{bmatrix}\mu\nu\\2\end{bmatrix} + \frac{\partial}{\partial x_3}\begin{bmatrix}\mu\nu\\3\end{bmatrix} - \frac{\partial}{\partial x_4}\begin{bmatrix}\mu\nu\\4\end{bmatrix},$$
$$(\tau = 1,2,3,4).$$

For μ = ν = 4, this gives,

$$\frac{\partial}{\partial x_1}\begin{bmatrix}44\\1\end{bmatrix} = \frac{1}{2}g_{44}\frac{\partial}{\partial x_1}\left(\frac{\partial g_{41}}{\partial x_4} + \frac{\partial g_{41}}{\partial x_4} - \frac{\partial g_{44}}{\partial x_1}\right) = -\frac{1}{2}\frac{\partial^2 g_{44}}{\partial x^2{}_1}.$$

And so on. Finally all four give for **the left-hand side of (53)**,

$$\frac{1}{2}\left(\frac{\partial^2 g_{44}}{\partial x_1{}^2} + \frac{\partial^2 g_{44}}{\partial x_2{}^2} - \frac{\partial^2 g_{44}}{\partial x_3{}^2}\right) = -\frac{\mathbf{1}}{\mathbf{2}}\boldsymbol{\nabla}^2 \boldsymbol{g}_{44}.$$

Thus (53) yields in the Newtonian approximation,[244]

$$\mathbf{(68)}\ \boldsymbol{\nabla}^2 \boldsymbol{g}_{44} = \boldsymbol{\kappa\rho}.$$

[Einstein wrote this in the following form: $\Delta g_{44} = \kappa\rho$]. This is equivalent to

Newton's law of gravitation, Poisson's equation.

In the manuscript of the 1916 paper somebody added the following words between the number of the equation (68) and the equation:

"Heim B.A". This could mean, came finally home to the Newtonian limit... but then what does B.A mean?[245] It is not Einstein's hand writing, and it is unknown who added this writing, and what does it mean.

By equations (67) and (68) the equation for the gravitational potential becomes,[246]

$$(68a)\quad -\frac{\kappa}{8\pi}\int\frac{\rho d\tau}{r}.$$

And Newton's theory gives,

$$-\frac{K}{c^2}\int\frac{\rho d\tau}{r}.$$

Thus, while K is the constant of gravitation,

$$\kappa = \frac{8\pi K}{c^2} = 1{,}87\cdot 10^{-27}.$$

*Newtonian limit in its full compatibility – from the field equations and from the geodesic equation.* The recovery of the Newtonian limit was related to the condition $\sqrt{-g} = 1$.

**2.17 Geometry and Experience**

Einstein arrived at Section §22 "Behavior of Rods and Clocks in Static Gravitational Field. Bending of light rays. Motion of the Perihelion of the Planetary Orbit".[247]

In the manuscript he added another element to this title, and he regretted in the last minute and deleted it: Redshift. Einstein wrote in the title in the manuscript,

"Verschiebung der Spektrallinien" (Shift of Spectral Lines)[248].

When imposing a condition of static gravitation fields, Einstein realized that other components of the $g_{\mu\nu}$ must differ from the values given in (4) $g_{\sigma\tau} = \mathrm{diag}(-1, -1, -1, +1)$, by small quantities of the first order.

(4) can be written as $g_{\sigma\tau} = \delta_{\sigma\tau}$, $g_{\sigma 4} = g_{4\sigma}$, $g_{44} = 1$. $\delta_{\sigma\tau}$ is equal to 1 or 0 when $\sigma = \tau$ or $\sigma \neq \tau$, respectively.  Hence for a field produced by a point mass (representing the sun) at the origin of the coordinates, Einstein obtained in November 18 as a first approximation for the metric field of the sun:[249]

$$(70)\qquad g_{\rho\sigma} = -\delta_{\rho\sigma} - \alpha\frac{x_\rho x_\sigma}{r^3},\ g_{44} = 1 - \frac{\alpha}{r}.$$

This was the weak fields approximation, the static gravitational field.

From (68a), $\alpha = 2KM/c^2$,

$$(70a)\ \alpha = \frac{\kappa M}{8\pi}.$$

In the manuscript of the 1916 paper Einstein added another term (the highlighted term below) to the right hand term above and deleted it,[250]

$$\alpha = \frac{\kappa M}{8\pi}\frac{\boldsymbol{M}}{\boldsymbol{A}}.$$

At the beginning of his paper, Einstein concluded after presenting the disk story that "In the general theory of relativity, space and time cannot be defined in such a way that spatial coordinate differences be directly measured by the unit measuring rod, and time by a standard clock".[251] Einstein ended his paper with an explanation of the disk story that is based on the metrical properties of space-time. Einstein's metric equation (3) demonstrated how a gravitational field changes a spatial dimension and a clock period.

Consider the metric equation presented in section §4,[252]

$$(3)\ ds^2 = g_{\mu\nu}dx_\mu dx_\nu$$

and a unit-measuring rod laid "parallel" to the x-axis. Then, $ds^2 = -1$; $dx_2 = dx_3 = dx_4 = 0$. Therefore, equation (3) gives, $-1 = g_{11}dx_1^2$.[253]

Suppose that this rod also lies *on* the x-axis. In this case, the first of equations (70) gives:[254]

$$g_{11} = -\left(1 + \frac{\alpha}{r}\right).$$

This equation and equation (3) in the form: $-1 = g_{11}dx^2_1$, yield,

$$(71)\ dx = 1 - \frac{\alpha}{2r}.$$

Einstein then arrived at the following conclusion, "The unit measuring rod therefore appears a little shortened with respect to the coordinate system by the presence of the gravitational field, if it is laid in the radial direction"[255].

As to the length of a measuring rod in the tangential direction: we set $ds^2 = -1$; $dx_1 = dx_3 = dx_4 = 0$; $x_2 = r$, $x_1 = x_3 = 0$. Therefore, equation (3) gives,[256]

(71a) $-1 = g_{22}dx_2^2 = -dx_2^2$.

With the tangential position, therefore, the gravitational field of the point mass has no influence on the length of a rod.

The initial motivation for presenting the rotating disk thought experiment in section §3 was to show that coordinates of space and time have no direct physical meaning, and Euclidean Geometry breaks down. Einstein used a coordinate dependent description and explanations from special relativity in order to convince the reader that this is indeed the case. Using equations (3) and (70) Einstein demonstrated,[257]

"Thus Euclidean geometry does not apply even to a first approximation in the gravitational field, if we wish to take one and the same rod, independently of its location and orientation, as a realization of the same interval. However, a glance at (70a) and (69) shows that the expected deviations are too small to be noticeable in measurements of the earth's surface".

Let us rewrite equation (71) in the following way (Einstein did not perform this step in the 1916 paper, but it is implicit in his derivation),

$$[(71b)]\ \boldsymbol{dx} = 1 - \frac{2KM}{c^2 2r} = \mathbf{1} + \frac{\boldsymbol{\Phi}}{\boldsymbol{c^2}}.$$

Let us come back for a moment to 1907. Einstein wrote in 1907, "By assuming this, we obtain a principle which, if it is true, has great heuristic meaning. For we obtained by theoretical consideration of the processes which take place relatively to a uniformly accelerating reference system (K'), information as to the course of processes in a homogeneous gravitational field (K)".[258]

Einstein finally extended his 1907 crude equivalence principle. His 1907 equivalence principle was embodied in the following equation, $(1 + \gamma\xi/c^2) = (1 + \Phi/c^2)$, the right term holds for a system *K'* and the left term for a system *K*.[259] These factors reappeared in almost every equation in the 1911 paper and generally, in the initial version of Einstein's coordinate-dependent general theory of relativity until 1912.

Recall that in section §3 of his 1912 paper, "the Speed of Light and the Statics of the Gravitational Fields", Einstein formulated his findings from his 1911 Prague paper and his 1907 review paper on the theory of relativity, "If we measure time in $S_1$ [lower gravitational potential] with a clock *U*, *we must measure the time in* $S_2$ [higher gravitational potential] *with a clock that goes 1 +* $\Phi/c^2$ *slower than the clock U if you compare it with the clock U in the same place*".[260]

In 1907 Einstein explained that,[261]

"In this sense we may say, that the process occurring in the clock – and more generally any physical process – proceeds faster the greater the gravitational potential at the position of the process taking place.

There are now 'clocks', which are located at different gravitational potentials and whose rates can be controlled very precisely; these are the producers of the spectral lines. From the above it is concluded that the light coming from the sun's surface, which is due to the pressure, is larger by about one part in two millionth of the wavelength than that of light produced by the same substance on earth".

The above description was the first time that Einstein described gravitational redshift.

In the 1916 review paper, starting from the line element (3), Einstein derived gravitational redshift. Einstein wrote, "Further, let us also examine the time coordinate examined by the rate of a unit clock, which is arranged at rest in a static gravitational field".[262]

For the clock period we set, ds = 1; $dx_1 = dx_2 = dx_3 = 0$. Thus, $1 = g_{44}dx_4^2$. Consider,[263]

$$\boldsymbol{dx_4} = \frac{1}{\sqrt{g_{44}}} = \frac{1}{\sqrt{1+(g_{44}-1)}} = \mathbf{1} - \frac{\boldsymbol{g_{44}} - \mathbf{1}}{\mathbf{2}}.$$

$$[1 - \frac{g_{44}-1}{2} = 1 + \frac{\Phi}{c^2}].$$

And according to (68a) we obtain,[264]

$$(72)\ \ dx_4 = 1 + \frac{\kappa}{8\pi}\int \frac{\rho d\tau}{r}.$$

Einstein concluded, "The clock goes then more slowly if it is placed near ponderable masses. It follows that the spectral lines of light reaching us from the surface of large stars must appear displaced towards the red end of the spectrum". Einstein added a footnote that according to Freundlich, spectroscopical observations on fixed stars of certain types indicate the existence of the effect of this kind. However, a final test of this consequence is still pending.[265]

Einstein's 1916 review article was written *after* Schwarzschild had found a complete exact solution to Einstein's field equations; a solution which satisfied the same conditions as the approximate solution (70); nevertheless, Einstein preferred in his 1916 paper to write his November 18, 1915 approximate solution. Why did not Einstein use the Schwarzschild solution? It appears that Einstein preferred his approximate procedure upon Schwarzschild exact solution (which contained a singularity at R = 0). The reason was now obvious: he could readily derive equation (71) using his approximate solution (70); and this led him to conclude that Euclidean geometry does not hold even to *first approximation* in the gravitational field. Einstein needed the first approximation solution to arrive at this conclusion. However, there was another reason for preferring (70) upon the Schwarzschild exact solution. Einstein obtained the deflection of a ray of light passing by the sun using his approximate scheme and Huygens principle.

Einstein then gave the explanation from his November 18, 1915 paper for bending of light rays in a gravitational field,[266] but expanded it, and fully derived the bending $2\alpha/\Delta$ from (73) and (70).

Einstein's starting point was his November 18 quadratic equation,[267]

$$(73)\ ds^2 = g_{\mu\nu} dx_\mu dx_\nu = 0.$$

(The world lines of light rays are geodesic null lines).

In special relativity $g_{\mu\nu} = \text{diag}(-1, -1, -1, +1)$, and thus,

$$-dx_1^2 - dx_2^2 - dx_3^2 + dx_4^2 = 0.$$

However, in general relativity, $g_{\mu\nu}$ are not equal to these constant values.

If the direction, the ratio $dx_1 : dx_2 : dx_3$, is given, equation (73) gives the quantities,[268]

$$\frac{dx_1}{dx_4}, \quad \frac{dx_2}{dx_4}, \quad \frac{dx_3}{dx_4}$$

and the velocity,

$$\sqrt{\left(\frac{dx_1}{dx_4}\right)^2 + \left(\frac{dx_2}{dx_4}\right)^2 + \left(\frac{dx_3}{dx_4}\right)^2} = \gamma,$$

is defined in the sense of Euclidean geometry.[269]

These are the equations that Einstein wrote before (equations [(I)] and [(II)]) in the manuscript.[270]

Einstein concluded that the path of the light rays must be curved with respect to the coordinate system, if the $g_{\mu\nu}$ are not constant. If n is the direction perpendicular to the propagation of light, then Huygens principle shows that the light ray [considered from the plane (γ, n)], has the curvature $-\frac{\partial\gamma}{\partial n}$.

Einstein examined the curvature undergone by a ray of light passing by a mass M (located at the origin of the coordinates) at the distance Δ from this mass. The ray travels parallel to the x axis of a coordinate system. The total curvature of the ray (expected to be positive if it is concave towards the origin) is given approximately by,[271]

$$\boldsymbol{B = \int_{-\infty}^{+\infty} \frac{\partial\gamma}{\partial x_1} dx_2}.$$

Using (73) and (70), Einstein wrote for γ the following,[272]

$$\boldsymbol{\gamma = \sqrt{-\frac{g_{44}}{g_{22}}} = 1 + \frac{a}{2r}\left(1 + \frac{x_2{}^2}{r^2}\right)},$$

where according to (70), $g_{44} = 1 - \frac{\alpha}{r}$, and $g_{22} = -1 - \alpha\frac{x_2{}^2}{r^3}$.

Inserting this value of γ into the equation of B and carrying out the calculation leads to the final result,[273]

$$(74)\ \mathrm{B} = \frac{2\alpha}{\Delta}.$$

And using equation (70a) and (69) for the value of α, this gives,[274]

$$(74)\ B = \frac{\kappa M}{4\pi\Delta}.$$

The above highlighted equations were written in the manuscript of the 1916 paper on page 44. Einstein cut this paper on the right hand side in a shape of a square for an unknown reason. Equation (74) was already starting page 45.[275]

Einstein concluded, "A ray of light passing by the sun therefore undergoes a deflection of 1.7", one passing the planet Jupiter gets an amount of 0.2".[276]

Einstein ended his paper by quoting the final equation from his November 18 paper, the equation for the perihelion advance in the sense of motion after a complete orbit,[277]

$$(75)\quad \varepsilon = 24\pi^3 \frac{a^2}{T^2 c^2 (1-e^2)},$$

where, *a* denotes the major semi-axis, *c* is the velocity of light, *e* is the eccentricity, and *T* the orbital period. Einstein wrote in a footnote,[278] "With respect to the calculation, I refer to the original treatises": Einstein's November 18 paper and Schwarzschild's 1916 paper.[279] Einstein then wrote, that calculation gives for the planet Mercury a precession value of 43" per century, corresponding exactly to astronomical observation.

Einstein thus ended his exposition of his theory with three experimental tests: the bending of light rays near the mass of the sun (not yet verified); gravitational red shift (Freundlich's observations had already indicated the existence of this effect, but further observations were supposed to verify this); and the precession of the perihelion of Mercury.[280] This is very typical to Einstein's papers: he does a theoretical analysis in the paper, and he ends it by proposing experimental tests; he cares about experiments.

I am indebted to Prof John Stachel for his assistance and invaluable suggestions. It should be noted that the contents of this paper are the sole responsibility of the author.

## Endnotes

[1] Einstein, 1907c, pp. 206-207.

[2] Ehrenfest, 1909, p. 918.

[3] Ehrenfest, 1909, p. 918.

[4] .Stachel, 1980 in Stachel (2002), p. 246

[5] Einstein, 1912b, p. 356.

[6] Ehrenfest, Paul, "Zur Frage der Entbehrlichkeit der Lichtäthers", *Physikalische Zeitschrift* 13 (1912), pp. 317-319.

[7] Einstein to Ehernfest, *CPAE*, vol 5, Doc. 409; Norton, 2004, p. 71; Stachel, 1982 in Stachel (2002), p. 184.

[8] Einstein, doc. 44, "Paul Ehrenfest in Memoriam", 1934, in Einstein, 1956, p. 238.

[9] Stachel, 1989 in Stachel (2002), pp. 323-324; Einstein to Ehernfest, December 26, 1915, *CPAE*, vol 8, Doc. 173; Einstein to Besso, January 3, 1916, *CPAE*, vol 8, Doc. 178.

[10] Einstein to Schlick, December 14, 1915, *CPAE*, Vol. 8, Doc 165; Stachel, 1989 in Stachel (2002), p. 323.

[11] Einstein, 2010.

[12] Einstein, 1914b.

[13] Einstein, 2010; Einstein, 1916.

[14] Einstein, 2010, p. 11; Einstein, 1916, p. 781.

[15] Lorentz, Einstein, and Minkowski, 1915, pp. 56-73.

[16] Einstein, 2010, p. 1.

[17] Einstein, 1916, p. 769.

[18] Einstein, 1916, p. 770.

[19] Einstein, 1916, p. 770.

[20] Einstein, 1916, p. 770.

[21] Einstein, 1916, p. 773.

[22] Seelig, 1956, p. 82; Seelig, 1954, p. 97.

[23] Einstein, 1916, p. 770.

[24] Einstein, 1916, p. 770.

[25] Newton, 1726/1997, p. 18.

[26] Einstein, 1914c, pp. 344-346.

[27] Einstein, 1914c, p. 344.

[28] Einstein, 1916, p. 771.

[29] Einstein, 1916, p. 771.

[30] Einstein, 1920b, p. 21, *CPAE*, Vol. 7, Doc. 31, p. 265.

[31] Einstein, 1905a, p. 891.

[32] Einstein, 1916, p. 771.

[33] Einstein, 1916, p. 772.

[34] Einstein, 1916, p. 772.

[35] Einstein, 1914b, p. 1032.

[36] Einstein, 1916, pp. 772-773.

[37] Einstein, 1916, p. 773.

[38] Einstein, 1907b, p. 461.

[39] Einstein, 1907b, p. 461.

[40] Einstein, 1911a, p. 898.

[41] Einstein, 1911a, p. 908.

[42] *CPAE*, "Einstein on Grossmann and Relativity: The Collaboration with Marcel Grossmann", Vol. 4, p. 295.

[43] Abraham, 1912b, p. 1056.

[44] Norton, 1993/2007, in Renn (2007), p. 429.

[45] Stachel, 1980 in Stachel (2002), pp. 246-247.

[46] Stachel, 1980, in Stachel (2002), p. 255.

[47] Einstein, 1916, p. 774.

[48] Einstein, 1916, p. 775.

[49] Einstein, 1916, p. 775.

[50] Einstein, 1916, p. 775.

[51] Einstein, 1912b, p. 356.

[52] Einstein, 1907b, p. 455.

[53] Einstein, 1907b, p. 457.

[54] Einstein, 1907b, p. 458.

[55] Einstein, 1911, p. 906.

[56] Einstein, 1916, p. 775.

[57] Einstein, 1916, p. 775.

[58] Einstein, 1916, p. 775.

[59] Einstein, 1917a, p. 53.

[60] Einstein, 1917a, p. 53.

[61] Einstein, 1917a, pp. 54-56.

[62] Stachel, 1980 in Stachel (2002), p. 248.

[63] Einstein, 1916, p. 776.

[64] Satchel, 2010, p. 17.

[65] Einstein, 1916, p. 776.

[66] Einstein, 1914b; Stachel, 1989, in (2002). The Point Coincidence Argument was extensively discussed in the literature; for instance, Howard, 1999 and Howard and Norton, 1993.

[67] Stachel, 1989 in Stachel (2002), p. 325.

[68] Stachel, 2010, p. 16.

Howard and Norton think that Einstein's immediate inspiration for the Point Coincidence Argument came from the paper by Erich Kretschmann, published before Einstein wrote to Ehrenfest. Their claim is that the paper, "Über die prinzipielle Bestimmbarkeit der berechtigten Bezugssysteme beliebiger Relativitätstheorien" ("On Determining the Principle of the Legitimate Reference System of any Theory of Relativity"), clearly anticipates essential elements of the Point Coincidence Argument. In a 1917 publication of his, Kretschmann cited this publication for further details when referring to Einstein's 1916 Point Coincidence Argument, and he even introduced the German term "Koinzidenzen" to replace another term he had used in 1915 in relation to his argument. Howard and Norton, 1993, p. 53. In addition, Kretschmann uses Einstein's 1916 public version of the Point Coincidence Argument to conclude that any theory could be put into a form satisfying Einstein's principle of general covariance. However, Stachel says that there is a difference between Einstein's private and public versions of the Point Coincidence Argument. In the public argument Einstein indeed did not provide a criterion for singling out theories. However, the private argument's criterion of general covariance does. Stachel, 2010, p. 19.

[69] Einstein to Ehernfest, December 26, 1915, *CPAE*, vol 8, Doc. 173; Einstein to Besso, January 3, 1916, *CPAE*, vol 8, Doc. 178.

[70] Einstein, 1916, p. 776.

[71] Einstein, 1905a, p. 893.

[72] Einstein, 1916, p. 774.

[73] Einstein, 1916, p. 774.

[74] Einstein, 1916, pp. 776-777.

[75] Stachel, 2010, pp. 17-18.

[76] Einstein, 2010, p. 7.

[77] Einstein, 1916, p. 777.

[78] Einstein, 1916, p. 777.

[79] Frank, 1949/1979, Einstein, Vorwort.

[80] Einstein, 1916, p. 777.

[81] Einstein, 1914b, pp. 1032-1033; for the equation from Einstein, 1914b see the first paper in the series of these papers, "From the Berlin 'Entwurf' Field Equation to the Einstein Tensor I: October 1914 until Beginning of November 1915. .

[82] Einstein, 1916, p. 778.

[83] Einstein, 1916, p. 778.

[84] Einstein, 1916, p. 779.

[85] Einstein, 1916, p. 779.

[86] Einstein, 1916, p. 779.

[87] Einstein, 1916, p. 779.

[88] Einstein, 1914b, p. 1033.

[89] Einstein, 1916, p. 780.

[90] Einstein, 1916, p. 781.

[91] Einstein, 2010, p. 11.

[92] Einstein, 1916, p. 781.

[93] Einstein, 1914b, pp. 1036-1037.

[94] Einstein, 1916, p. 782.

[95] Einstein, 1916, p. 782.

[96] Einstein, 1916, p. 783.

[97] Einstein, 1914b, p. 1037.

[98] Einstein, 1916, p. 783.

[99] Einstein, 1916, p. 784.

[100] Einstein, 1914b, p. 1038.

[101] Einstein, 1916, p. 784.

[102] Einstein, 1916, p. 786.

[103] Einstein, 1914b, p. 1039.

[104] Einstein, 1916, p. 788.

[105] Einstein, 1914b, p. 1040; Einstein, 1916, p. 788.

[106] Einstein, 1914b, p. 1040; Einstein, 1916, p. 788.

[107] Einstein, 1915a, p. 779; for the equation fom Einstein's November 1915 paper see the second paper in the series: see the first paper in the series of these papers, "From the Berlin 'Entwurf' Field Equation to the Einstein Tensor II: November 1915 until March 1916.

[108] Einstein, 1915a, p. 779.

[109] Einstein, 1914b, pp. 1041-1042; Einstein, 1916, p. 789.

[110] Einstein, 1911a, p. 906.

[111] Einstein, 1916, p. 789.

[112] Stachel, private conversation.

[113] Einstein, 1916, p. 790.

[114] Einstein, 1914b, pp. 1045-1046.

[115] Einstein, 1916, p. 791.

[116] Einstein, 1916, p. 792.

[117] Einstein, 1916, p. 792.

[118] Einstein, 1914b, pp. 1047-1048.

[119] Einstein, 1916, p. 792.

[120] Einstein, 1916, p. 793.

[121] Einstein, 1916, p. 793.

[122] Einstein, 1916, p. 795.

[123] Einstein, 2010, p. 22.

[124] Einstein, 1914b, p. 1050.

[125] Einstein, 1916, p. 795.

[126] Einstein, 1916, p. 796.

[127] Einstein, 2010, p. 23.

[128] Einstein, 1916, p. 796.

[129] Einstein, 1916, p. 796.

[130] Einstein, 2010, p. 23 and the additional piece of paper attached to it.

[131] Einstein, 2010, p. 23 and the additional piece of paper attached to it.

[132] Einstein, 1914b, p. 1051.

[133] Einstein, 1914b, p. 1051.

[134] Einstein, 1914b, p. 1051.

[135] Einstein, 1914b, p. 1051.

[136] Einstein, 1916, p. 797.

[137] Einstein, 1915a, p. 780.

[138] Einstein, 1916, p. 798.

[139] Einstein, 1916, p. 798.

[140] Einstein, 1916, p. 799.

[141] Einstein, 1914b, p. 1053.

[142] Einstein, 1914b, pp. 1053-1054.

[143] Einstein, 1916, p. 799.

[144] Einstein, 1914b, p. 1053.

[145] Einstein, 1916, p. 800; Einstein, 1914b, p. 1053.

[146] Einstein, 1916, p. 800; Einstein, 1914b, p. 1053.

[147] Einstein, 1914b, p. 1053.

[148] Einstein, 1916, pp. 800-801.

[149] Einstein, 1916, p. 801.

[150] Einstein, 2010, p. 27.

[151] Einstein, 1915a, p. 780.

[152] Einstein, 1916, p. 801.

[153] Einstein, 1916, p. 802; Einstein, 1915a, p. 783.

[154] Einstein, 1916, p. 802; Einstein, 1915a, p. 783.

[155] Einstein, 1916, pp. 801-802.

[156] Einstein, 1916, p. 802.

[157] Einstein, 1916, p. 802.

[158] Einstein, 1916, pp. 802-803.

[159] Einstein, 1916, p. 812.

[160] Einstein, 1916, p. 803; Einstein, 1915c, p. 832.

[161] Einstein, 1915d, p. 845.

[162] Einstein, 1916, p. 803.

[163] Einstein, 1916, p. 803.

[164] Einstein, 1916, p. 803.

[165] Einstein, 1916, p. 803.

[166] Einstein, 1916, p. 804.

[167] Einstein, 1916, p. 804.

[168] Einstein, 2010, p. 30.

[169] Einstein, 1915a, p. 785.

[170] Einstein, 1915a, p. 784.

[171] Einstein, 1916, p. 804.

[172] Einstein, 1916, pp. 804-805.

[173] Einstein, 1916, p. 805.

[174] Einstein, 1915a, p. 784.

[175] Einstein, 1915a, p. 784.

[176] Einstein, 1915a, p. 784.

[177] Einstein, 1916, p. 805; Einstein, 1915a, p. 784.

[178] Einstein, 1916, p. 805; Einstein, 1915a, p. 784.

[179] Einstein, 2010, p. 31.

[180] Einstein, 1916, p. 805; Einstein, 1915a, p. 784.

[181] Einstein, 1916, p. 806; Einstein, 1915a, p. 785.

[182] Einstein, 2010, p. 31.

[183] Einstein, 1916, p. 806.

[184] Einstein, 1916, p. 806.

[185] Einstein, 1915a, p. 785; written here with Einstein's summation convention, but did not appear in this way in the November 4 paper.

[186] Einstein, 1915a, p. 785.

[187] Einstein, 1915a, p. 785.

[188] Einstein, 1916, p. 806.

[189] Einstein, 1916, p. 806.

[190] Einstein, 1916, p. 806.

[191] Einstein, 2010, p. 32.

[192] Einstein, 1916, p. 806.

[193] Einstein, 1916, p. 807.

[194] Einstein, 1915a, p. 785.

[195] Einstein, 1916, p. 807.

[196] Einstein to Ehrenfest, January 17, 1916, *CPAE*, Vol. 8, Doc. 182.

[197] Einstein to Ehrenfest, January 24 or later, 1916, *CPAE*, Vol. 8, Doc. 185.

[198] Einstein to Ehrenfest, January 24 or later, 1916, *CPAE*, Vol. 8, Doc. 185.

[199] Einstein to Ehrenfest, January 17, 1916, *CPAE*, Vol. 8, Doc. 182.

[200] Einstein to Ehrenfest, January 24 or later, 1916, *CPAE*, Vol. 8, Doc. 185.

[201] Einstein, 1916, p. 808.

[202] Einstein, 1922/1956, pp. 84-86.

[203] Einstein, 1916, p. 808.

[204] Einstein, 1916, p. 809.

[205] Einstein, 1916, p. 809.

[206] Einstein, 1916, p. 809.

[207] Einstein, 1916, p. 810.

[208] Einstein, 1916, p. 799.

[209] Einstein, 1916, p. 810.

[210] Einstein, 1916, p. 809.

[211] Einstein, 1916, p. 811.

[212] Einstein, 1914, p. 1060.

[213] Einstein, 1916, p. 811.

[214] Einstein, 1914, p. 1062.

[215] Einstein, 1916, pp. 797; 812.

[216] Einstein, 1916, p. 812.

[217] Einstein, 1916, p. 812.

[218] Einstein, 1916, p. 813.

[219] Einstein, 1916, p. 798.

[220] Einstein, 1916, p. 813.

[221] Einstein, 1916, p. 813.

[222] Einstein, 1916, p. 814.

[223] Einstein, 1916, p. 815.

[224] Einstein, 1915b, p. 799.

[225] Einstein, 2010, p. 40a.

[226] Einstein, 1916, p. 815.

[227] Einstein, 1916, p. 789.

[228] Einstein, 1916, p. 815.

[229] Einstein, 1916, p. 816.

[230] Einstein, 2010, p. 40.

[231] Einstein, 2010, p. 40.

[232] Einstein, 2010, p. 41.

[233] Einstein, 2010, p. 41.

[234] Einstein, 1916, p. 816.

[235] Einstein, 1915c, p. 835.

[236] Einstein, 1915c, p. 834.

[237] Einstein, 1915c, p. 835.

[238] Einstein, 1915c, p. 835.

[239] Einstein, 1916, p. 817.

[240] Einstein, 2010, p. 41.

[241] Einstein, 1916, p. 817.

[242] Einstein, 1916, p. 817.

[243] Einstein, 1916, p. 818.

---

[244] Einstein, 1916, p. 818.

[245] Einstein, 2010, p. 42.

[246] Einstein, 1916, p. 818.

[247] Einstein, 1916, p. 818.

[248] Einstein, 2010, p. 42.

[249] Einstein, 1916, p. 819; Einstein, 1915c, p. 833.

[250] Einstein, 2010, p. 43.

[251] Einstein, 1916, p. 775.

[252] Einstein, 1916, p. 819.

[253] Einstein, 1916, p. 819.

[254] Einstein, 1916, p. 819.

[255] Einstein, 1916, p. 820.

[256] Einstein, 1916, p. 820.

[257] Einstein, 1916, p. 820.

[258] Einstein, 1911, p. 900.

[259] Einstein, 1907b, p. 461.

[260] Einstein, 1911a, p. 906.

[261] Einstein, 1907b, pp. 458-459.

[262] Einstein, 1916, p. 820.

[263] Einstein, 1916, p. 820.

[264] Einstein, 1916, p. 820.

[265] Einstein, 1916, p. 820.

[266] Einstein, 1915c, p. 834.

[267] Einstein, 1916, p. 821.

[268] Einstein, 1916, p. 821.

[269] Einstein, 1916, p. 821.

[270] Einstein, 2010, p. 40.

[271] Einstein, 1916, p. 821.

[272] Einstein, 1916, p. 822.

[273] Einstein, 1916, p. 822.

[274] Einstein, 1916, p. 822.

---

[275] Einstein, 2010, p. pp 44-45.

[276] Einstein, 1916, p. 822.

[277] Einstein, 1916, p. 822; Einstein, 1915c, p. 839.

[278] Einstein, 1916, p. 822.

[279] Schwarzschild, 1916a.

[280] Einstein, 1916, p. 822.

---